\documentclass[twocolumn,showpacs,pra,floatfix,superscriptaddress]{revtex4-1}
\usepackage[utf8]{inputenc}
\usepackage{graphicx}
\graphicspath{{.}{./figures/}}
\usepackage[english]{babel}
\usepackage{amsmath,amssymb}
\usepackage[usenames,dvipsnames]{color}
\usepackage{bbold}
\usepackage{times}
\usepackage[colorlinks,bookmarks=false,citecolor=blue,linkcolor=red,urlcolor=blue]{hyperref}
\usepackage[normalem]{ulem}

\newcommand*{\paraup}{\ensuremath{\protect\substack{\bullet \bullet \\ \circ \circ}} }
\newcommand*{\paradown}{\ensuremath{\protect\substack{\circ \circ \\ \bullet \bullet}} }
\newcommand*{\diagmain}{\ensuremath{\substack{\bullet \circ \\ \circ \bullet}} }

\newcommand{\bra}[1]{\langle #1|}
\newcommand{\ket}[1]{| #1\rangle}
\newcommand{\mboti}[1]{{\mbox{\tiny{#1}}}}
\newcommand{\mbt}[1]{{\mbox{\tiny{#1}}}}

\begin{document}
\title{Propagation and jamming dynamics in Heisenberg spin ladders}

\author{Carlo B. Krimphoff}
\affiliation{Institut f\"ur Theoretische Physik, Leopold-Franzens Universit\"at Innsbruck, A-6020 Innsbruck, Austria}

\author{Masudul Haque}
\affiliation{Max-Planck-Institut f{\"u}r Physik komplexer Systeme, 01187 Dresden, Germany}

\affiliation{Department of Mathematical Physics, National University of Ireland, Maynooth, Ireland}

\author{Andreas M. L{\"a}uchli}
\affiliation{Institut f\"ur Theoretische Physik, Leopold-Franzens Universit\"at Innsbruck, A-6020 Innsbruck, Austria}

\begin{abstract}
  We investigate the propagation dynamics of initially localized excitations in
  spin-$\frac 12$ Heisenberg ladders.  We consider initial states with two
  overturned spins, either on neighboring sites on the same leg or on
  the two sites of a single rung, in an otherwise polarized
  (ferromagnetic) background.  Compared to the corresponding dynamics
  in a chain (single leg), we observe several additional modes of
  propagation.  We connect these propagation modes to features of the
  spectrum of the ladder system, and to different effective models
  corresponding to different segments of the spectrum.  In addition to
  the regular propagation modes, we observe for one mode a peculiar
  `jamming' dynamics where components of the excitations remain
  localized in an unusual manner.  A comparison with the spin-1
  bilinear-biquadratic chain is developed and explored, where a
  similar phenomenon is shown to occur.

\end{abstract}

\date{\today}

\maketitle

%\bibliographystyle{apsrev4-1_custom}

%\nocite{*}
\section{Introduction}
During the last few years, interest in coherent unitary dynamics of quantum many-body systems has
grown rapidly \cite{Eisert2015,Polkovnikov2011,Dziarmaga2010}, motivated primarily by the possibility of tracking
non-dissipative dynamics in real time in ultracold atomic systems \cite{LewensteinBook2012,Bloch2012,Bloch2007,Greiner2002a,Kinoshita2006,Hofferberth2007}.
An emerging theme is the propagation and binding dynamics of spatially localized objects in quantum
lattice systems \cite{Winkler2006, Fukuhara2013a, Fukuhara2013b, Ganahl2012a, Konno2005, GanahlBowling2013, Wollert2012, Vlijm2015, Haque2010, Sharma2014, Alba2013, Petrosyan2007, Petrosyan2008, Jin2008, Khomeriki2010a, Deuchert2012, Santos2012}.
In particular, after an experiment with interacting bosonic atoms,
which highlighted the interaction-induced longevity of repulsive pairs \cite{Winkler2006},
much attention has focused on the \emph{binding} (and anti-binding) of localized excitations and the
dynamics of these bound clusters, both in itinerant systems
\cite{Petrosyan2007, Petrosyan2008, Jin2008, Khomeriki2010a, Deuchert2012, Santos2012} and in spin chains
\cite{Ganahl2012a, GanahlBowling2013, Wollert2012, Haque2010, Sharma2014, Alba2013, Vlijm2015}.
In a lattice spin system, one can consider a few downturned spins in a (ferromagnetic) background of
up-spins as particles in a background of empty sites. These initially localized magnons can bind to each
other due to interactions.  In recent experiments with optical lattices that realize spin chains,
such localized objects have been created on single sites and on a pair of neighbouring sites, and
the propagation of single magnons as well as of bound magnon pairs has been explicitly tracked in
real time \cite{Fukuhara2013a, Fukuhara2013b,Fukuhara2015}.  Trapped ion systems have recently been applied to
study spin dynamics and spectral properties of the XY chain with long range spin exchange
\cite{Jurcevic2014,Richerme2014,Jurcevic2015}.  

The setup of the experiments \cite{Fukuhara2013a, Fukuhara2013b,Fukuhara2015} is well-suited to explore
multiple-chain situations or a square lattice.  Motivated by this
experimental capability, in this work we consider dynamics on a two-leg
Heisenberg spin-$\tfrac{1}{2}$ ladder.  We start with simple initial product states with two neighboring overturned
spins, either on the same leg or on the same rung, and analyze the subsequent dynamics.  The
Heisenberg ladder can be regarded as the minimal extension of a spin chain toward
two-dimensionality.  Nevertheless, we find a rich sequence of new behaviors, including three
ballistic modes of motion and a peculiar non-ballistic mode.  We relate the propagation dynamics to 
spectral decompositions of the initial states. We rely on numerically exact real-time evolutions using a Krylov-space 
technique on the one hand, and analytical considerations mostly based on mappings to sectors of simpler spin chains
on the other hand.

%%%%%% FIGURE %%%%%% FIGURE %%%%%% FIGURE %%%%%% FIGURE %%%%%% FIGURE %%%%%%%%%%
\begin{figure}
  \centering
  \includegraphics[width=0.5\textwidth]{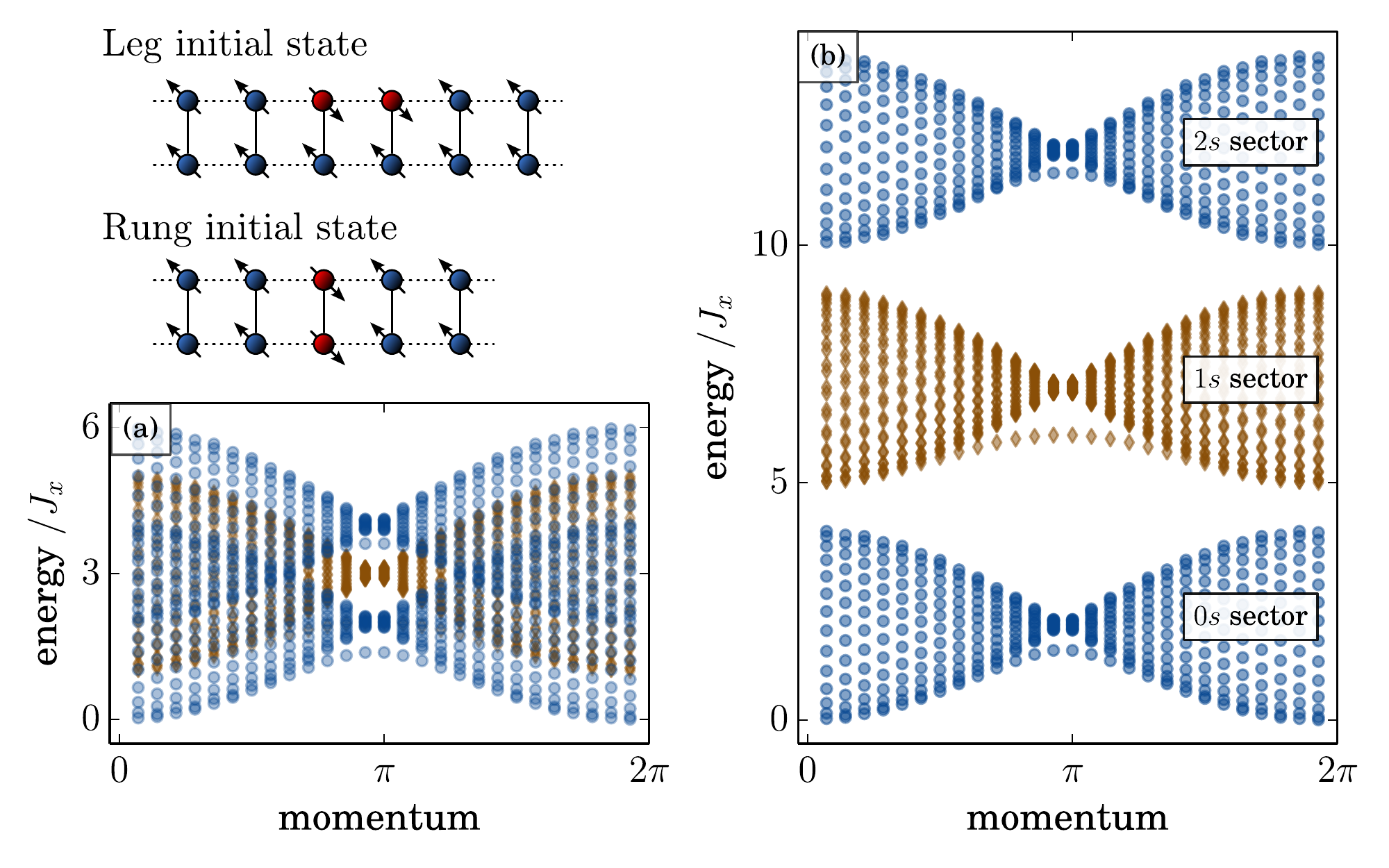}
  \caption{
    (Top left) Schematic of the initial states.
    The leg initial state has two excitations on the same leg while the rung initial state has both excitations on the same rung of the ladder.
(a) Energy spectrum of the two-leg Heisenberg ladder with two
excitations in $2L=54$ lattice sites, for a coupling strength $\chi =
1$, and (b) at larger coupling $\chi = 5$.
The lower band ($0s$ sector) is equivalent to a spin-$1$ Heisenberg chain, while the upper band ($2s$ sector) constitutes a spin-$\frac 12$ $\Delta=\frac 12$ XXZ chain. 
The center band ($1s$ sector) corresponds to two spin-$\frac 12$ XXZ chains, one with $\Delta=1$
and one with $\Delta=0$.  
The eigenstates in the top ($2s$) and bottom ($0s$) bands are symmetric with respect to leg exchange;
eigenstates in the center band ($1s$) are antisymmetric.
}
\label{fig:intro}
\end{figure}
%%%%%% FIGURE %%%%%% FIGURE %%%%%% FIGURE %%%%%% FIGURE %%%%%% FIGURE %%%%%%%%%%

The corresponding simpler situation in the one-dimensional anisotropic Heisenberg spin chain (XXZ
spin chain) is by now well understood.  A single $\downarrow$ spin $(S^z=-\tfrac{1}{2})$ in a
ferromagnetic $\uparrow$ $(S^z=+\tfrac{1}{2})$ background propagates as a free particle and the
magnetization as a function of space and time is given by a Bessel function
\cite{Konno2005,Fukuhara2013a}.  The density of states as a function of the velocity 
has a maximum at the maximum group velocity; hence a dominant wave front propagates with this particular velocity
\cite{Ganahl2012a}.  With multiple $\downarrow$ spins, the $S^z$-anisotropy $\Delta$
plays the role of interactions.  Two neighboring $\downarrow$ spins are strongly bound in the large
anisotropy (Ising) limit, and propagate as a slow particle.  For $\Delta < 1/\sqrt{2}$, no binding is
observed and each excitation propagates independently like a single magnon.  At intermediate
$\Delta$, both propagation modes are seen simultaneously
\cite{Ganahl2012a,Fukuhara2013b}.  In this work we present the much richer
phenomenology of corresponding situations in the two-leg ladder.

The system Hamiltonian is given by
\begin{align}
    H = -J_x \sum \limits_{y=1}^2 \sum \limits_{x=1}^{L} {\bf S}_{x,y} \cdot {\bf S}_{x+1,y}
    - J_y \sum \limits_{x=1}^{L} {\bf S}_{x,1} \cdot {\bf S}_{x,2} \,,
    \label{eq:Hamiltonian}
\end{align}
with periodic boundary conditions in the $x$-direction, i.\,e. $x=L+1$
is identified with $x=1$.  The relative coupling strength is denoted
by $\chi = J_y/J_x$.  We consider the ferromagnetic ladder ($J_{x,y}>0$), but a simultaneous change of signs of both couplings ($J_{x,y}<0$) does not affect the dynamics since all initial states we consider are time reversal invariant.
Note that we focus on Heisenberg interactions ($\Delta=1$) and equal signs of the couplings ($\chi\geq 0$), as
appropriate for the experimental platform \cite{Fukuhara2013a,
  Fukuhara2013b}.  The local Hilbert space of a particular rung can be 
described by the usual basis of three triplet states and one singlet state: $\ket{t_{\pm}}$,
$\ket{t_0}$, and $\ket{s}$.  The rung initial state
\begin{align}
    \ket{\mbox{rung}} = S^-_{L/2,1} S^-_{L/2,2} \ket{0} \,,
    \label{eq:RungInit}
\end{align}
corresponds to a single $\ket{t_-}$ embedded in a chain of $\ket{t_+}$ rungs. The leg initial state
\begin{align}
    \ket{\mbox{leg}} = S^-_{L/2,1} S^-_{L/2+1,1} \ket{0}
    \label{eq:LegInit}
\end{align}
has two down spins on neighboring rungs: it has
components with two $\ket{s}$'s, one $\ket{s}$ and one $\ket{t_0}$, and two $\ket{t_0}$'s.  We will
refer to the sub Hilbert spaces spanned by these states as the $2s$, $1s$, and $0s$ sectors.

The boundary conditions do not play a role in the dynamics until a signal from the initial positions
reaches the boundary.
Periodic boundary conditions in the $x$-direction allow us to use the $x$-momentum as a conserved quantum number.  The Hamiltonian is also symmetric under
exchange of legs, so that each eigenstate has well-defined parity under leg exchange.

The rung magnetization $\langle S^z_{x,1} + S^z_{x,2} \rangle$, viewed as a function of
position $x$ and time $t$, displays the propagation phenomena.
By using exact diagonalization, we investigate this observable under time evolution with (\ref{eq:Hamiltonian}) for both initial states. 
The propagation generally happens as a collection
of well-defined wave fronts.
The time evolution of various projection operators, like the one projecting on two neighboring flipped spins on a rung of the ladder
\begin{align}
    P^{\downarrow}_i = \langle (S^z_{i,1} - 1/2)(S^z_{i,2} - 1/2) \rangle \,,
    \label{eq:ProjDef}
\end{align}
are useful to study the propagation of quasiparticles, such as bound states.

Our goal is to explain the wave fronts (and in one case a non-ballistically propagating mode) observed for the real time evolution
in terms of the structure of the Hilbert space.

The text is structured in the following way:
In Section~\ref{sec:Overview}, we present an overview of the Hilbert
space and the energy spectrum.
Sec.~\ref{sec:spectrum} is devoted to the various subspaces and mappings of the two-leg Heisenberg ladder.
We briefly outline all dynamical effects and observed modes of propagation in Sec.~\ref{sec:Rich}.
A quantitative discussion of the leg initial state is given by Sec.~\ref{sec:MainLeg}, followed by the detailed discussion of the rung initial state in Sec.~\ref{sec:MainRung}.
In Sec.~\ref{sec:RungAnalysis}, we focus on the connection of real space observation and spectral decompositions
while in Sec.~\ref{sec:current}, we investigate the rung initial state via different current operators leading us to the interpretation of a specific mode of propagation as jamming dynamics.
In Sec.~\ref{sec:BLBQ}, we present an explanation of non-ballistic dynamics observed for the rung initial state by means of perturbation theory.

\section{Overview of Spectrum and propagation modes}\label{sec:Overview}
%%%%% TABLE %%%%% TABLE %%%%% TABLE %%%%% TABLE %%%%% TABLE %%%%% TABLE %%%%%
\begin{table*}%[!htbp]
\centering
\begin{tabular}{c|c|c}
\toprule
\multicolumn{2}{c|}{symmetric subspace} & antisymmetric subspace $({\bf 1s})$\\
\hline
$\chi \ll 1 $ & $\chi \gg 1$ & any $\chi$ \\
\colrule
$\begin{array}{c} \mbox{\textbf{2 particles per leg}} \\ \rightarrow \mbox{spin-} \frac 12, \Delta=1 \mbox{ chain} \end{array}$ &
    $\begin{array}{c} {\bf 0s} \\ \rightarrow \mbox{spin-}1, \Delta=1 \mbox{ chain} \end{array}$ &
    $\begin{array}{c} \mbox{\textbf{2 particles per leg}} \\ \rightarrow \mbox{spin-}\frac 12, \Delta=1 \mbox{ chain} \end{array}$ \\

    $\begin{array}{c} \mbox{\textbf{1 particle per leg}} \\ \rightarrow \mbox{non-interacting BH chain} \end{array}$ &
    $\begin{array}{c} {\bf 2s} \\ \rightarrow \mbox{spin-}\frac 12, \Delta=\frac 12 \mbox{ chain} \end{array}$ &
    $\begin{array}{c} \mbox{\textbf{1 particle per leg}} \\ \rightarrow \mbox{spin-}\frac 12, \Delta=0 \mbox{ chain} \end{array}$ \\
\botrule
%\toprule
\end{tabular}
  \caption{Overview of the structure of the Hilbert space for the two-leg ladder with two
    excitations.  The bold titles indicate separations of the Hilbert space appropriate for each
    case.  The chain systems to which a subspace is mapped are listed with arrows.  The
    antisymmetric subspace ($1s$, right column) is disjoint from the rest of the Hilbert space,
    irrespectible of the coupling stength $\chi$.  The symmetric subspace is divided according to
    the number of excitations per leg for small $\chi$ (left column) and according to the singlet
    number for $\chi \gg 1$ (center column).  }
  \label{tab:HilbertSpaces}
\end{table*}
%%%%% TABLE %%%%% TABLE %%%%% TABLE %%%%% TABLE %%%%% TABLE %%%%% TABLE %%%%%

In this Section, we first outline how different parts of the Hilbert space and energy spectrum can
be mapped onto simpler models (\ref{sec:spectrum}).  These mappings will allow us to identify the
physical content of the different propagation modes.  In \ref{sec:spectra_and_dynamics} we comment
on how spectral features dictate real-time propagation dynamics, in particular, how the speed of
bound excitations is determined by the dispersion.  In \ref{sec:Rich}, we provide an overview of the
propagation modes observed with our two initial states.

\subsection{Spectrum of the two-particle sector and mappings to chain Hamiltonians}\label{sec:spectrum}
The spectrum of the ladder with two particles and at large $\chi$ is shown in Figure
\ref{fig:intro}(b).  There are three `bands' of width $4J_x$ spaced at distance $J_y={\chi}J_x$. (At
small $\chi$ the bands overlap and are not visually distinguishable, see Fig.~\ref{fig:intro}a.)
Each band resembles the spectrum of two particles in a single chain, consisting of a `continuum'
part shaped like a bow-tie and a 'bound state' part showing up as a single-line dispersion under
each band.  The `continuum' eigenstates are dominated by configurations with the two spins separated
from each other.  In the eigenstates comprising the `bound state' branch, there is strong
probability for the two spins to neighbor each other.

The top and bottom bands contain states that are symmetric under leg exchange ($k_y=0$).  The center
band is antisymmetric ($k_y=\pi$).  The continuum states in the center band are nearly (two-fold)
degenerate.  In total there are $L^2$ symmetric states and $L^2-L$ antisymmetric states, which sum
up to $ {2L}\choose {2}$ states.  
Since the Hilbert space grows quadratically with the number of sites, the Hamiltonian can easily be diagonalized numerically for system sizes up to several hundred rungs.

At large $\chi$, there is an energetic separation between the three energy sectors, because the
energy is dominated by the rung coupling $J_y$ and fewer rung singlets (more rung triplets) are
energetically favored.  The bottom, middle and top bands correspond respectively to $0s$, $1s$ and
$2s$ sectors.
Although the different bands overlap in energy for smaller values of $\chi$, conservation of $k_y$
does not allow for matrix elements between symmetric and antisymmetric states.

The lowest band ($0s$ sector) corresponds to the physics of a spin-1 chain, as all rungs of the ladder
remain in a $S=1$ triplet state. At large $\chi$, the $0s$
part of the symmetric sector can be mapped onto the spin-1 bilinear-biquadratic chain (BLBQ):
\begin{align*}
    H_\mboti{blbq} = -J_\mboti{bl} \sum \limits_{\langle i,j\rangle} {\bf T}_i \cdot {\bf T}_{j}
    -J_\mboti{bq} \sum \limits_{\langle i,j\rangle} \left( {\bf T}_i \cdot {\bf T}_{j} \right)^2 
\end{align*}
with the couplings $J_\mboti{bl} = J_x/2$ and $J_\mboti{bq} = J_x^2/(8 J_y) = J_x/8\chi$, and
${\bf T}_i$ is a spin-1 operator associated with the rung $i$.  The mapping is detailed in Appendix
\ref{sup:BBMmap}.  The two terms are the leading and subleading terms in an expansion in $\chi$.  A
mapping to the simpler spin-$1$ chain $(J_{\mboti{bq}}=0)$ is obtained at first order as shown in
\cite{Mila1998}, and revised as well in App.~\ref{sup:BBMmap}.

The middle band ($1s$ sector) eigenstates correspond to the physics of two spin-$\frac 12$
anisotropic Heisenberg (XXZ) chains.  In App.~\ref{sup:AntiMap}, we show that the antisymmetric
subspace with two particles is exactly mapped onto a combination of two spin-$\frac 12$
chain Hamiltonians of the form
\begin{align}
  \begin{split}
    H_A =& \, -\frac{J_x}{2} \sum_{\langle i,j\rangle} \left\{ {\bf S}_i^+ {\bf S}_j^- + \mbox{h. c.} \right\} \\
    & - \Delta_A J_x \sum_{\langle i,j\rangle} {\bf S}_i^z {\bf S}_j^z - h_A \sum_i {\bf S}^z_i + \epsilon_A \,.
  \end{split}
  \label{eq:H12gen}
\end{align}
We use the symbol ${\bf S}_i$ to denote spin-$\frac 12$ operators, as before; the single site index
indicates that this Hamiltonian lives on a chain rather than a ladder.  One Hamiltonian corresponds
to $\Delta_A=0$, $h_A = J_x + \frac{1}{2}J_y$, and $\epsilon_A = 0$, i.\,e. a so-called XX chain
Hamiltonian.  The other corresponds to $\Delta_A=1$, $h_A = (J_x + J_y)/2$, and $\epsilon_A = J_x$,
i.\,e. the $SU(2)$-symmetric Heisenberg Hamiltonian.  The $\Delta_A=0$ Hamiltonian maps onto a chain
of non-interacting fermions; neighboring magnons do not interact and no binding phenomenon is
observed.  The spectrum of this subsector contains only a continuum part.  The bound magnon branch
visible in the middle band of the spectrum is associated with the $\Delta_A=1$ Hamiltonian.
Regardless of the coupling strength $\chi$, the XX and the Heisenberg subspace are completely
separated due to the conservation of parity of the number of particles per leg.  These subspaces
define an integrable subsystem which is also found for ladders with more than two legs
(App.~\ref{sup:AntiMap}).   

The top band ($2s$ sector) at large $\chi$ maps onto another spin-$\frac 12$ XXZ chain Hamiltonian,
this time with $\Delta=\frac 12$.  This mapping has appeared in the literature previously
\cite{Mila1998, Chaboussant1998, Giamarchi1999}, and is outlined in App.~\ref{sup:MapD12}.

The different pictures and mappings discussed so far will turn out to be useful for the discussion
of dynamics in the strong coupling regime of the ladder $\chi \gg 1$.  To complete the picture, we
consider the opposite limit of small coupling $\chi \ll 1$.  In this limit, the separation into $0s$,
$1s$ and $2s$ is not appropriate.  Similarly to the discussion of the antisymmetrized subspace
(App.~\ref{sup:AntiMap}), we consider a separation of the symmetrized subspace into configurations
with an even particle number per leg $\mathcal{B}_2$ and configurations with exactly one particle
per leg $\mathcal{B}_1$.  These two Hilbert spaces can be mapped (App.~\ref{sup:LowChiMap})
respectively to a spin-$\frac 12$ Heisenberg chain with two excitations ($\mathcal{B}_2$) and a
non-interacting Bose Hubbard chain (BH) carrying two excitations ($\mathcal{B}_1$).  

The structure of the Hilbert space is summarized in Table~\ref{tab:HilbertSpaces}.

\subsection{Energy spectra and real space dynamics} \label{sec:spectra_and_dynamics}
Real space dynamics of closed systems is completely encoded in the spectral properties of the
initial state.  In addition to the time evolution, we numerically compute spectral decompositions,
i.\,e.~the overlap of the particular initial states with the eigenstates of the systems
(App.~\ref{sup:SpecFcts}).  

In some cases, the spectrum directly gives the speed of a mode of propogation.  For initial states
showing a weight distribution by means of a well-defined and sufficiently smooth energy momentum
relation $\epsilon_k$, the group velocity is defined as the derivative $v_g = \partial \epsilon_k /
\partial k$.  If the dispersion $\epsilon_k$ has an inflection point at $k=k^*$, then the time
evolution is expected to display the propagation of a wave front expanding with a velocity $v$ given
by the group velocity at the inflection point
\begin{align}
    \left. v = \frac{\partial \epsilon_k}{\partial k} \right|_{k=k^*} \,.
    \label{eq:DosArg}
\end{align}
A similar argument has been used in Ref.~\cite{Ganahl2012a} in their discussion of
bound magnon pair propagation in the spin-$\frac 12$ XXZ chain.  We provide further details in
App.~\ref{sup:EnginDyn}.

\subsection{Overview of propagation modes} \label{sec:Rich}
We summarize here the propagation phenomena observed for the leg and the rung initial states; they
will be treated in greater depth in subsequent Sections.

\begin{figure}
  \centering
  \includegraphics[width=\linewidth]{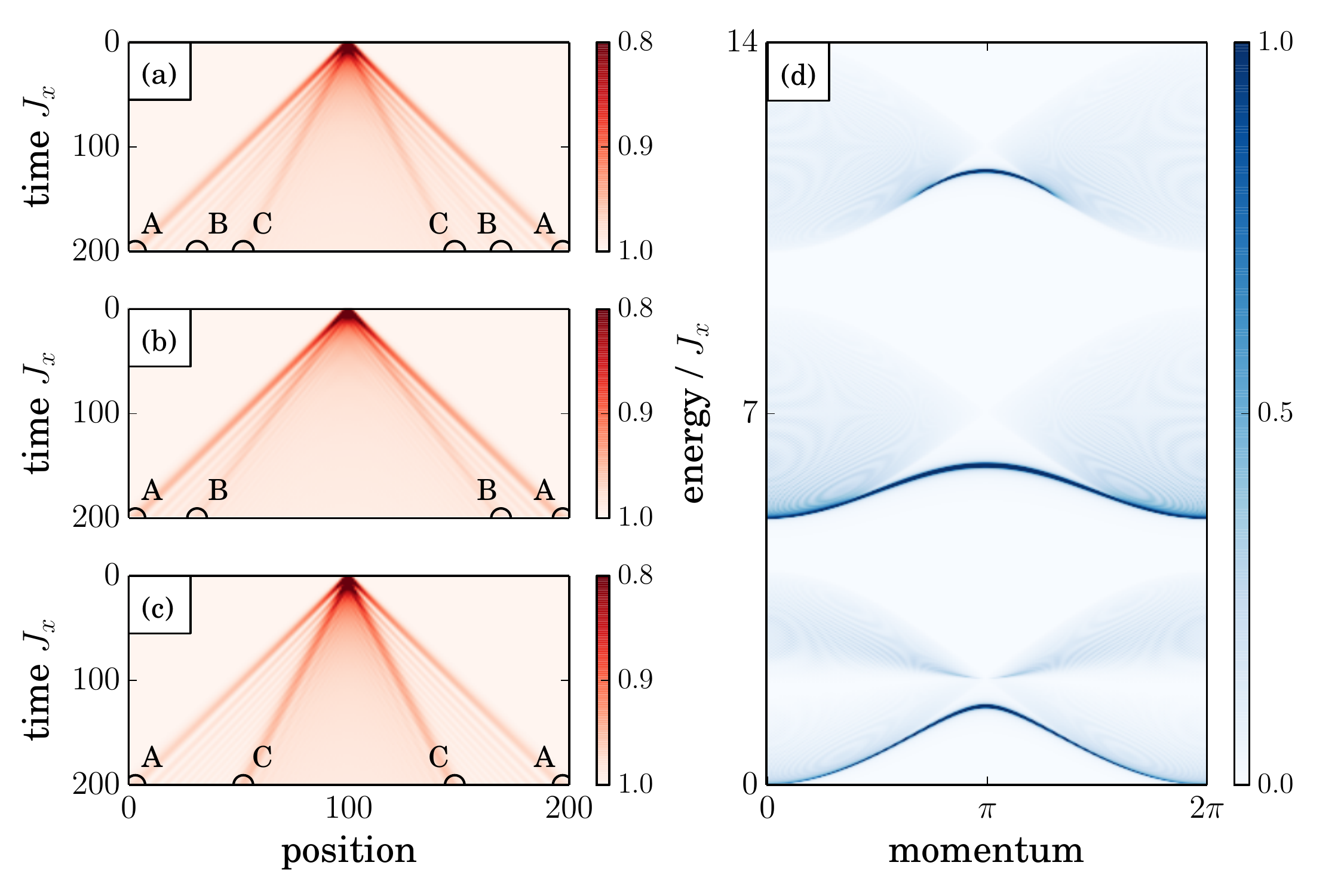} 
  \caption{ Real-time dynamics (a,b,c) and spectral decomposition (d) for the leg initial state at a
    coupling strength $\chi=5$.  (a) Time evolution of the local magnetization $\langle
    S^z_{i,1} + S^z_{i,2} \rangle$.  (b,c) Dynamics after decomposition into symmetrized 
      (b) and antisymmetrized (c) part of the leg initial state.  The position of each
    wave front at $t = 200/J_x$ is marked with half-circle symbols.
}
 \label{fig:LegReal}
\end{figure}

Starting from the leg initial state, we observe three distinguishable wave fronts propagating
ballistically with different velocities, Fig.~\ref{fig:LegReal}a.  The velocities of the fastest (denoted A)
and the slowest (denoted C) modes are independent of the coupling strength.  For the fastest mode this is not
suprising as it corresponds to the propagation of two single excitations~\cite{Fukuhara2013a}.  The
slowest propagation mode can be seen as corresponding to the bound-magnon or two-string mode known in the
spin-$\frac 12$ Heisenberg chain \cite{Ganahl2012a, Fukuhara2013b}.  The fact that this mode has a velocity
independent of $J_y$ follows from a symmetry and is therefore not a trivial effect.  The velocity of
the intermediate velocity wave front (denoted B) shows a moderate dependence on the coupling strength as its velocity changes
monotonically within a closely bounded interval of velocities, Fig.~\ref{fig:TripVelo}.  This
intermediate-velocity mode crosses over from the bound state of a spin-$\frac 12$ Heisenberg chain 
in the limit $\chi\ll 1$ to the bound state of two excitations in a spin-$1$ Heisenberg chain in the limit $\chi\gg1$.  

\begin{figure*}
  \centering
  \includegraphics[width=1\textwidth]{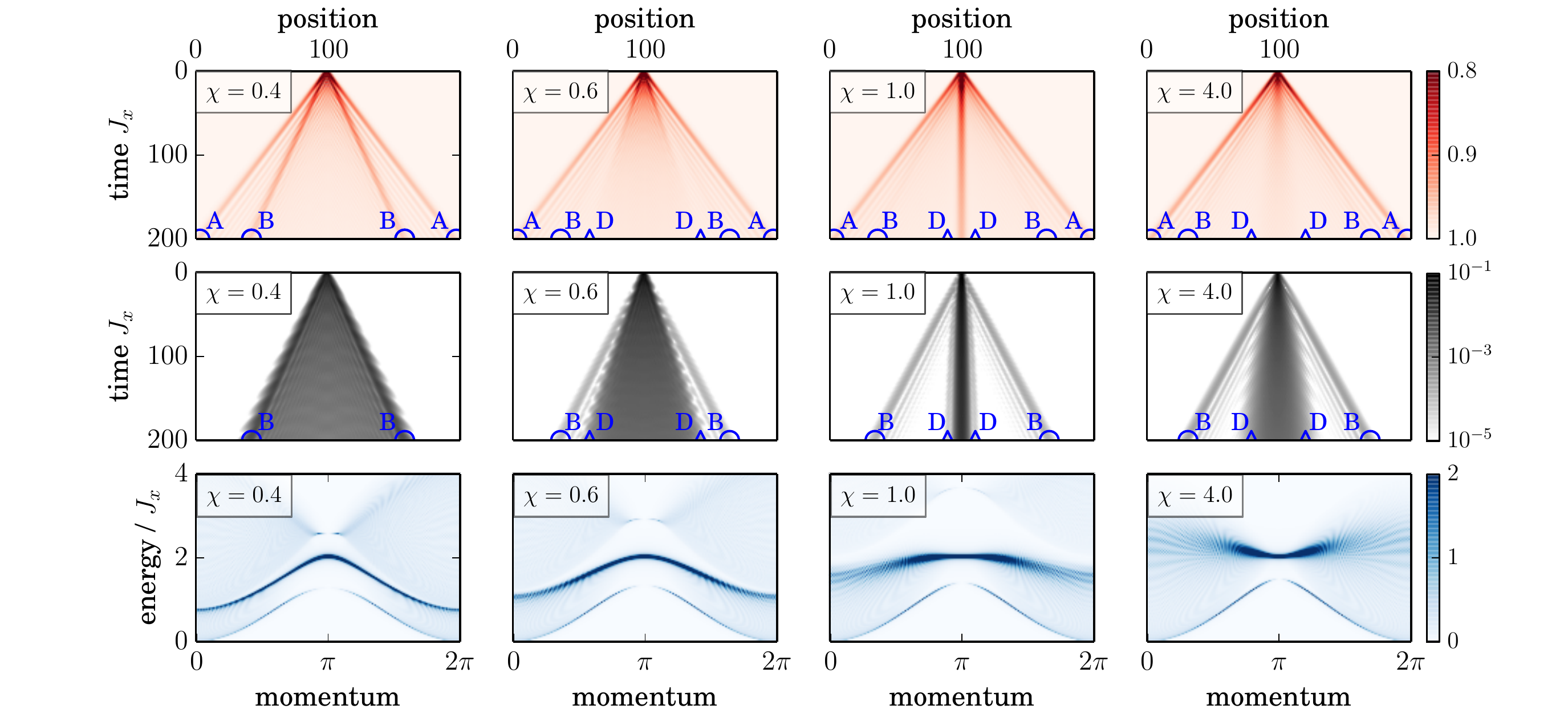}
  \caption{Time evolution and spectral decomposition of the rung initial state.
    We depict the magnetization $\langle S^z_{i,1}(t) + S^z_{i,2} \rangle$ in the first row, the projection on rung states $\langle P^{\downarrow}_i(t) \rangle$ (see Eq.~\eqref{eq:ProjDef} for a definition) in the second row and the lowest band of the corresponding spectral decompositions of the initial state in the third row.
  The coupling strength varies from $\chi = 0.4, 0.6, 1.0, 4.0$ from column one through four.
  Each wave front is marked with symbols at $t = 200 / J_x$.
  While waves fronts marked with A and B have the same physical interpretation as A and B in Fig.~\ref{fig:LegReal}, the mode of propagation marked with D is exclusively found for the rung initial state.
  }
    \label{fig:MainRung}
\end{figure*}

In contrast to the leg initial state, dynamics of the rung initial state is strongly sensitive to
the coupling $\chi$, Fig.~\ref{fig:MainRung}.  Two of the three modes identified for the leg initial
state --- the single particle mode (A) and the bound triplet mode (B) --- are also found in the rung
case, although with a reduced intensity for mode (B).  The spin-$\frac 12$ Heisenberg chain bound magnon
state (the slowest mode (C) for the leg initial state) does not appear for the rung initial state.
In addition, there is a novel mode of propagation (denoted D) existing exclusively for the rung
initial state which shows qualitatively different behavior and displays a peculiar effect we refer
to as jamming.  This mode of propagation expands ballistically for small $\chi$ on the observed time
scales.  The corresponding wave fronts however decay exponentially instead of algebraically
indicating propagation of resonances instead of bound states.  Beyond a threshold value of $\chi$,
ballistic propagation via wave fronts of this particular mode is lost, Fig.~\ref{fig:WFT}.  Using
the ansatz $\sigma^2 = D t^\alpha$ for the width $\sigma$ (spatial standard deviation) from the
normalized density profile $P_i^\downarrow(t)$, we find that the exponent $\alpha$ has a
non-monotonic behavior, with a minimum near $\chi\approx1$, and saturating around $\alpha\approx1$
at large $\chi$, i.\,e. showing a diffusion-like spreading of the signal at large $\chi$.  We
characterize this diffusive-like mode using different current functions and observe another peculiar
effect: the movement of occupied rungs is inverted, i.\,e. occupied rungs in the right half of the
ladder move to the left.  This leads to the interpretation that propagation is slowed down by some
jamming mechanism resulting from the counterpropagation of different quasiparticles.

%%%%%% FIGURE %%%%%% FIGURE %%%%%% FIGURE %%%%%% FIGURE %%%%%% FIGURE %%%%%%%%%%
%%%%%% FIGURE %%%%%% FIGURE %%%%%% FIGURE %%%%%% FIGURE %%%%%% FIGURE %%%%%%%%%%

%%%%%% FIGURE %%%%%% FIGURE %%%%%% FIGURE %%%%%% FIGURE %%%%%% FIGURE %%%%%%%%%%
%%%%%% FIGURE %%%%%% FIGURE %%%%%% FIGURE %%%%%% FIGURE %%%%%% FIGURE %%%%%%%%%%

\section{Leg Initial state}\label{sec:MainLeg}
In this Section, we examine in detail the propagation modes that occur 
 for the leg initial state.

We observe three distinct propagating wave fronts, as shown in
Fig.~\ref{fig:LegReal}.  In the Subsections below, we analyze these
modes and interpret them through spectral features and mappings to
simpler chains.

\subsection{Single particle mode of propagation}\label{sec:SingleParticle}
The fastest propagation mode visible in the magnetization profile $\langle S^z_{i, 1}(t) + S^z_{i,
  2}(t) \rangle$ (Fig.~\ref{fig:LegReal}a, marked with letter ``A'') propagates with velocity $v = J_x$, and is independent of
the inter-chain coupling $J_y$.  This mode of propagation is ubiquitous in our spin ladder as well
as in spin chains for a wide variety of initial conditions, and corresponds to the propagation of
single-magnon excitations.  For example, it can be identified in the two-particle sector of an XXZ
spin chain as well \cite{Ganahl2012a, Fukuhara2013a, Fukuhara2013b}.

Figs.\ \ref{fig:LegReal}b and \ref{fig:LegReal}c show that the fastest mode (single-particle mode)
is present in both the symmetric and antisymmetric sectors of the Hilbert space.  All the energy
sectors (0s, 1s, 2s) contribute to this propagation mode.  The single-particle mode appears because
the initial state has spectral weight in the bow-tie-shaped continuum parts of the spectrum (in
addition to the weights in bound state branches which lead to more complicated modes).

\subsection{Spin-$\frac 12$ magnon bound states}
The slowest mode of propagation (denoted C in Fig.~\ref{fig:LegReal}) does not depend on the coupling strength. This suggests a relation
to the $1s$ sector which is also independent of $J_y$, App.~\ref{sup:AntiMap}.  Indeed, the leg
initial state is a linear combination of $2s$, $1s$ and $0s$ eigenstates and it is the antisymmetrized
$k_y=\pi$ part ($1s$ sector) which is responsible for the slowest mode of propagation.

We introduce a shorthand pictorial notation where $\bullet$ denotes a down spin and $\circ$ denotes
an up spin.  An example configuration of the ladder is
\begin{align*}
    \ket{\diagmain} = \ket{\diagmain}_{i} = S^-_{i,2} S^-_{i+1,1} \ket{0} \,,
\end{align*}
with the rest of the ladder (omitted in the notation) understood to be
up ($\circ$) spins.  
A decomposition of the leg initial state into symmetrized and antisymmetrized parts is given by
\[
\ket{\paraup} =  \frac{1}{2}(\, (\ket{\paraup} + \ket{\paradown} ) + ( \ket{\paraup} -
\ket{\paradown})\,)  ,
\] 
with $(\ket{\paraup} + \ket{\paradown}) \in \{ 0s\, \cup\, 2s\}$ and
$(\ket{\paraup} - \ket{\paradown}) \in 1s$.

By preparing the antisymmetrized part alone as the initial state and
observing the time evolution, Fig.~\ref{fig:LegReal}c, we verify
that the slow $J_y$-independent mode of propagation for the leg
initial state is inherited from the $1s$ sector.

As the antisymmetrized subspace is mapped to the spin-$\frac 12$ chain and the leg initial state
dictates an even number parity of excitations on a leg, we translate the initial state into two
overturned spins in a spin-$\frac 12$ Heisenberg chain.  The mode of propagation corresponds to two
excitations moving as a spin-$\frac 12$ magnon bound state and the legs of the ladder are
effectively decoupled.  For a spin-$\frac 12$ chain, these propagating bound objects have been
discussed in the literature utilizing the integrability (via Bethe ansatz) of the spin-$\frac 12$
chain ~\cite{Ganahl2012a}, and are refered to as propagating ``strings''.  An analogous
bound state mode is found for ladders with an arbitrary but even number of legs by constructing
initial states generalizing the form shown in Eq.~\eqref{eq:AntiMap2} (App.~\ref{sup:AntiMap}) to
ladders with more legs.

\subsection{Bound triplet mode of propagation} \label{sec:TripMode}
%%%%%% FIGURE %%%%%% FIGURE %%%%%% FIGURE %%%%%% FIGURE %%%%%% FIGURE %%%%%%%%%%
\begin{figure}
  \centering
  \includegraphics[width=\linewidth]{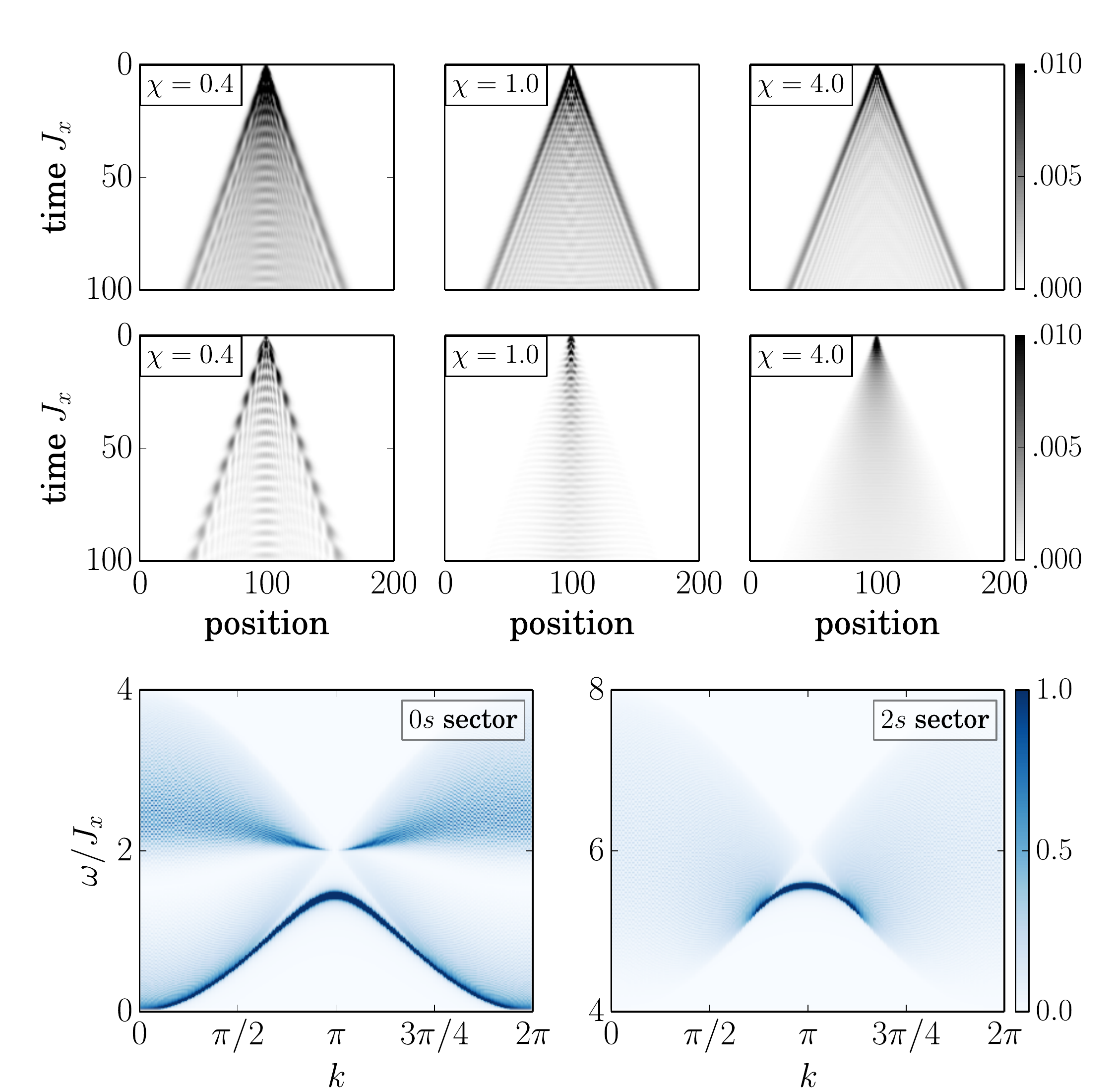}
  \caption{
Dynamics and spectral decomposition of the symmetrized leg initial state, 
$\frac{1}{\sqrt{2}}(\ket{\paraup} +\ket{\paradown})$, belonging to the $0s \cup 2s$ subspace.  Top
two rows: dynamics at different coupling strengths $\chi=J_y/J_x$.  Top row:  projector  
on neighboring triplets $P^{t_0}_i(t) = |\langle t_0, t_0 \ket{\psi(t)} |^2$ while $\ket{\psi(t)}$ is the time evolved symmetrized leg initial state.
Second row:
    projector on adjacent
    singlets $P^{s}_i(t) = |\langle s, s \ket{\psi(t)}|^2$. Here, $\ket{t_0, t_0} = \ket{t_0}_{i}\,
    \ket{t_0}_{i+1}$ and $\ket{s, s} = \ket{s}_{i}\,  \ket{s}_{i+1}$. 
Bottom row: spectral decomposition at $\chi=2$, shown in two panels (note the different values on the vertical axes). The $1s$ sector has no
overlap with the symmetrized initial state.  There is significant weight in the bound state
branches of both $0s$ and $2s$ sectors. 
  }
  \label{fig:SymLegDyn}
\end{figure}
%%%%%% FIGURE %%%%%% FIGURE %%%%%% FIGURE %%%%%% FIGURE %%%%%% FIGURE %%%%%%%%%%

\begin{figure}
  \centering
    \includegraphics[width=0.48\textwidth]{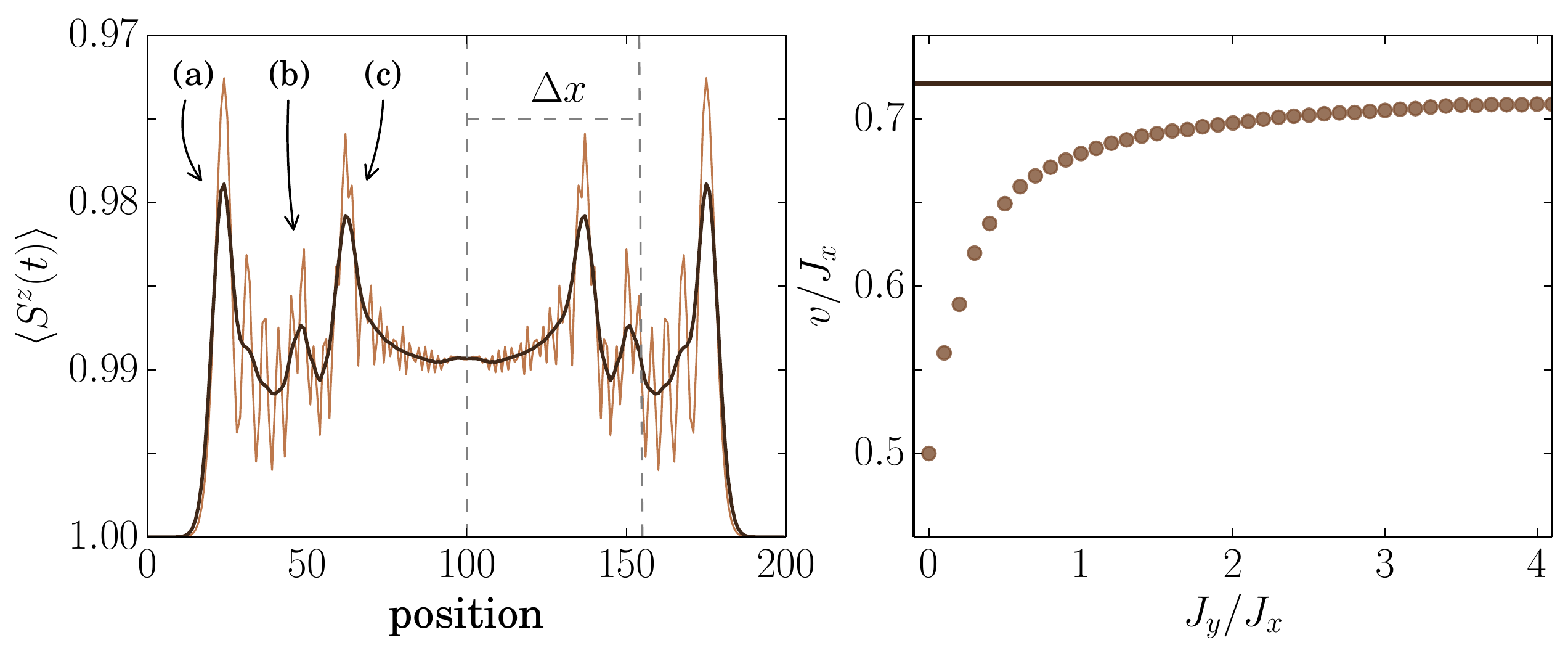}
    \caption{ Dynamics of the leg initial state.  Left: Snapshot of the local magnetization at $t =
      80/J_x$ after preparing the system as the leg state and letting it evolve with an intermediate
      coupling strength of $\chi = 1$.  We note the propagation of three different wave fronts,
      identified as: (a) single particle dynamics, (b) bound triplet mode of propagation and (c) two-body
      spin-$\frac 12$ bound state (two-string).  Raw data is given by the lighter, more noisy line
      while the smoothed profile is useful to extract an average expansion velocity and distinguish
      the different modes of propagation more clearly (black line).  Right: Average velocity
      ($\Delta x / t$) of the inflection point of the bound triplet mode (b) as a function of $\chi$: For
      decoupled legs $(J_y=0)$, the mode coincides with the propagation of a two-string.  For strong
      leg coupling, the system recovers the behavior of two neighboring $S^z=0$ sites in a $S^z= 1$
      polarized spin-1 chain with velocity $v \approx 1.44 J_{ \mbox{\tiny{bl} } }$ and
      $J_{\mbox{\tiny{bl}}} = J_x/2$.  }
    \label{fig:TripVelo}
\end{figure}

The third mode observed for the leg initial state (denoted B in Fig.~\ref{fig:LegReal}), with speed intermediate between the other two
modes, belongs to the symmetric part (0s~$\cup$~2s) of the Hilbert space, as seen from the
comparison of the symmetric and antisymmetric projections in Fig.~\ref{fig:LegReal}b,c. 
There is a weak dependence of the speed of this wave front on $\chi=J_y/J_x$.  For zero coupling,
the mode coincides with the spin-$\frac 12$ magnon bound state (speed $v=\frac{1}{2}J_x$), but with
increasing coupling the mode approaches a larger speed ($v\approx0.72J_x$),
Fig.~\ref{fig:TripVelo}.

By comparing the time evolution of the projection operators projecting on adjacent singlets
$\ket{s}$ and the operators projecting on neighboring triplets $\ket{t_0}$
(Fig.~\ref{fig:SymLegDyn}), we see that the wave front is almost always (apart from small $\chi$)
found for the triplet projection.  The symmetrized part of the leg initial state thus 
propagates as bound triplet pairs.  We will refer to this as the bound triplet mode.  

The bottom row of Fig.~\ref{fig:SymLegDyn} shows the spectral decomposition of the symmetrized part
of the leg initial state.  The $1s$ sector is not present in this decomposition, because this sector is
antisymmetric under leg exchange.  For both $0s$ and $2s$ sectors, the bound state branches are very
strongly populated.  The $2s$ sector does not contribute to a visible propagating bound state mode
(Subsection \ref{sec:2sdynamics}); the bound triplet mode arises from the bound state branch of the 0s
sector.  Except for small $\chi$, the physics (and form of the spectrum) of the $0s$ sector is
captured by the spin-1 chain with biquadratic interactions (BLBQ chain), App.~\ref{sup:BBMmap}.  For the
spin-1 chain, the energetically separated lower branch of the spectrum is mainly composed of bound
states of two neighboring $T^z=0$ sites in a polarized $T^z= 1$ background.  This is consistent with
our identification of this mode as a propagating bound state of two $\ket{t_0}$'s.   

At $\chi=0$, when the legs are decoupled, the leg initial state has the dynamics of a
spin-$\frac{1}{2}$ chain ($\Delta=1$) starting with two neighboring particles.  As a result this
mode coincides with the magnon bound state mode at $\chi=0$.  This is also why the symmetric leg
state dynamics has significant weight in the $\ket{s}_{i}\ket{s}_{i+1}$ projection for small $\chi$,
Fig.~\ref{fig:SymLegDyn}.  Spectrally, the $0s$ state is then mixed with the $2s$ state.  As the
coupling is increased, the $0s$ space gets transformed from a spin-$\frac 12$ chain structure to the
spin-$1$ chain.

The crossover of the symmetric sector mode from spin-$\frac{1}{2}$ chain physics (bound magnons) to
spin-$1$ chain physics (bound triplets) is demonstrated through the speed of the mode, studied as a
function of $\chi$ (Fig.~\ref{fig:TripVelo}).  The speed can be obtained from the magnetization
profiles (Fig.~\ref{fig:TripVelo} left) by smoothing the data (moving average filtering~\footnote{The moving average filtering is performed with a 5-point moving average, applied twice.
Our results do not depend noticeably on the smoothing procedure.})
and fitting the position of the relevant wave front to
$x_{\text{wave front}}-\frac{L}{2} = vt$.  (The wave front position is defined by the inflection point
of the magnetization profile, as in \cite{Bonnes2014}.) The speed of the bound triplet mode is seen to
increase smoothly from the value of the magnon bound state speed ($v=\frac{1}{2}J_x$) in the
spin-$\frac{1}{2}$ chain to the value of the $T^z=0$ bound state speed ($v\approx0.72J_x$) in the
spin-$1$ chain.  The value for the spin-$1$ chain is not analytically known (since the bound-state
spectral branch is not analytically expressible to the best of our knowledge), but the numerical
value $\approx0.72J_x$ is consistent with the value $\approx1.44J_{\text{bl}}$ we obtain by
numerically simulating the spin-1 chain.

%%%%%% FIGURE %%%%%% FIGURE %%%%%% FIGURE %%%%%% FIGURE %%%%%% FIGURE %%%%%%%%%%
%%%%%% FIGURE %%%%%% FIGURE %%%%%% FIGURE %%%%%% FIGURE %%%%%% FIGURE %%%%%%%%%%

\subsection{Absence of visible wave fronts for the $2s$ sector}\label{sec:2sdynamics}
Although both the leg and rung initial states have spectral weight in the topmost energy band ($2s$
sector), there is no prominent wave front associated explicitly with this band for finite $\chi$.  

The $2s$ sector maps to the spin-$\frac 12$ XXZ chain with $S^z$-anisotropy $\Delta=\frac 12$.  For
the spin-$\frac 12$ XXZ chain, a minimum anisotropy of $\Delta_c = 1/\sqrt{2}$ is required to observe
propagating two-body bound states~\cite{Ganahl2012a}.  Once $\Delta$ is decreased below
$\Delta_c$, the inflection point of the bound-state (two-string) dispersion is no longer present as
that part of the bound-state branch has merged into the continuum.  The disappearing of a
propagating bound pair at smaller $\Delta$ is discussed in greater detail in Appendix
\ref{sup:no_bound_state_at_small_Delta}.

\section{Rung Initial State}\label{sec:MainRung}
%%%%%% FIGURE %%%%%% FIGURE %%%%%% FIGURE %%%%%% FIGURE %%%%%% FIGURE %%%%%%%%%%
\begin{figure}
  \centering
  \includegraphics[width=0.495\textwidth]{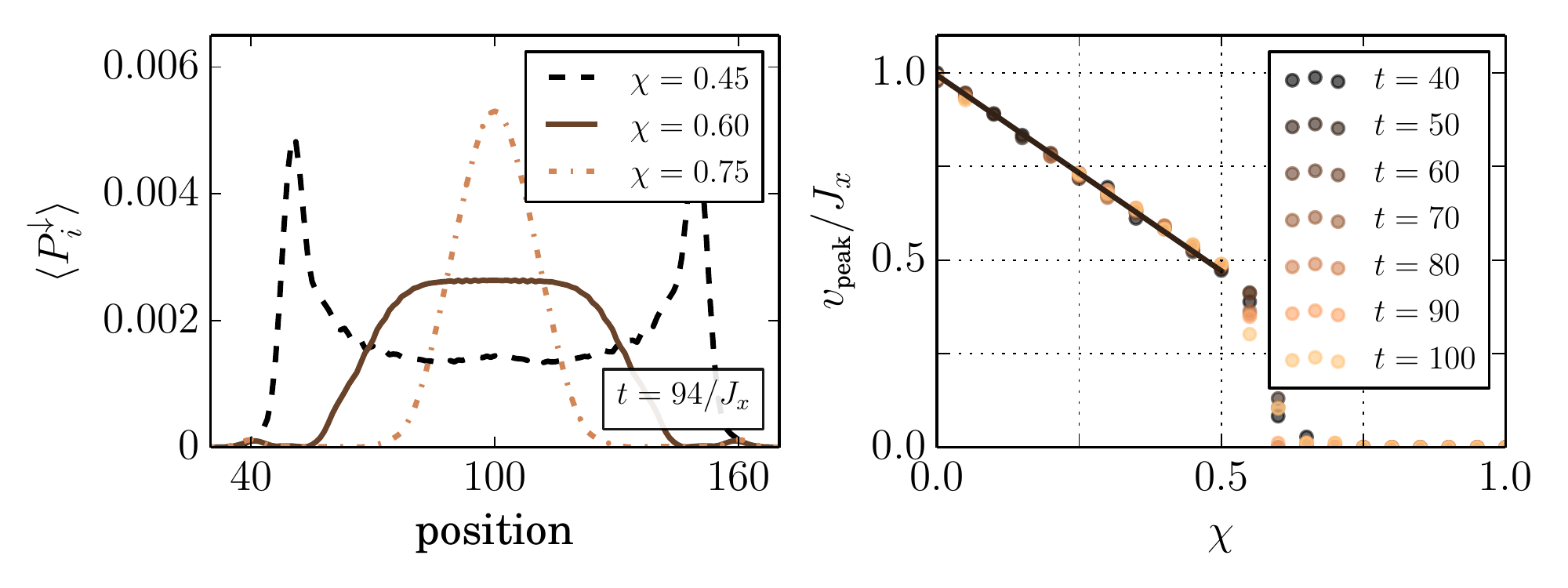}
  \caption{Crossover from ballistic to non-ballistic dynamics of the rung initial state.  Left:
    Snapshot of the smoothed rung density profile (projection on rung states) after propagation time
    $t=94 J_x$, for different coupling strengths.  Right: Average speed obtained from the peak
    position of the $P^\downarrow_i(t)$ profile.  Around $\chi\approx 0.5$, this `speed' drops to
    zero, indicating that there are no wave fronts but a plateau-like profile at larger $\chi$. }
  \label{fig:WFT}
\end{figure}
%%%%%% FIGURE %%%%%% FIGURE %%%%%% FIGURE %%%%%% FIGURE %%%%%% FIGURE %%%%%%%%%%

The rung initial state is symmetric under leg exchange ($k_y=0$) and hence has overlap only with the
$0s$ and $2s$ sectors.  The magnon bound pair mode (mode C), associated with the $1s$ band,
therefore does not appear for this initial state.  The other two propagation modes seen with the leg
initial state, the single-particle mode (mode A) and the bound triplet mode (mode B), are both
present.  However, the bound triplet signal is often too weak to be visible in the top row of
Fig.~\ref{fig:MainRung}, although it can be seen in the spin up rung projector (middle row).  We
will not discuss these further; in the following we focus on the rung-specific mode.  We first
describe the temporal dynamics of the magnetization (\ref{sec:RungAnalysis}), then connect it to the
spectral decomposition (\ref{sec:Rung_Spectral}), and then provide an alternate view of the
peculiarities of this mode using the dynamics of current-like observables (\ref{sec:current}).  We
also provide a perturbative analysis of some of the features (\ref{sec:BLBQ}).

Except for small $\chi$, the dynamics is dominated by the $0s$ sector, hence the dynamics with the
rung initial state closely resembles the dynamics of a spin-$1$ BLBQ chain.  We will exploit this
analogy in \ref{sec:current} and \ref{sec:BLBQ}.

%%%%%% FIGURE %%%%%% FIGURE %%%%%% FIGURE %%%%%% FIGURE %%%%%% FIGURE %%%%%%%%%%
\begin{figure}
  \centering
  \includegraphics[width=0.53\textwidth]{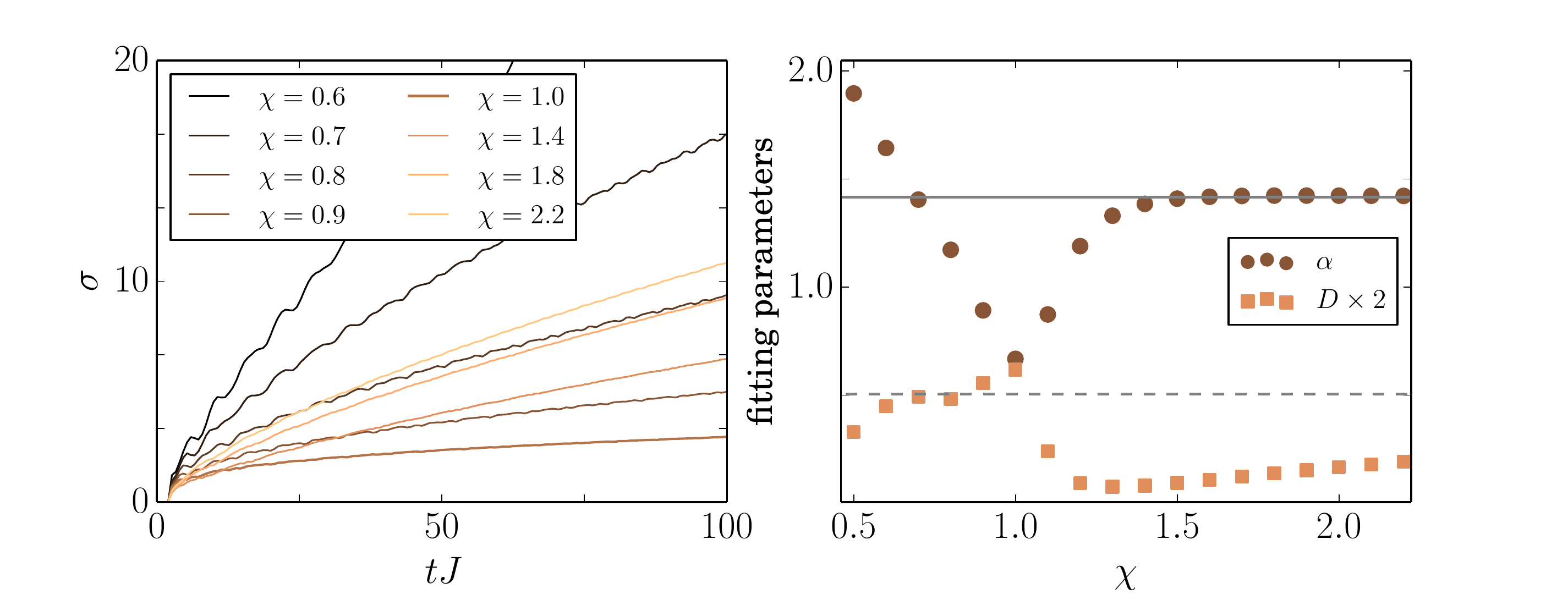}
  \caption{Characterizing non-ballistic spreading in the mode specific to the rung initial state. 
% $2L = 602$ lattice sites.
      Left: Time evolution of the width $\sigma(t)$ of normalized spatial profile
of  $P^\downarrow_i(t)$, as defined in Eq.\ \eqref{eq:sigma_def}.
  Right: Fitting coefficients from a fit to $\sigma^2(t) = D t^\alpha$.
  For $\chi \gg 1$, the behavior of $\sigma(t)$ is described by the
  corresponding quantity of the spin-1 chain (horizontal lines).  The spreading is slower than
  ballistic ($\alpha<2$) but faster than diffusive ($\alpha>1$).
}
  \label{fig:Sigma}
\end{figure}
%%%%%% FIGURE %%%%%% FIGURE %%%%%% FIGURE %%%%%% FIGURE %%%%%% FIGURE %%%%%%%%%%

\subsection{The rung-specific mode: non-ballistic dynamics} \label{sec:RungAnalysis}

For zero coupling, the rung initial state factorizes into two chains with a single overturned spin
in each.  Hence, dynamics for $\chi=0$ is given by fast single particle propagation.

As seen in Fig.\ \ref{fig:MainRung}, an increase of the ladder coupling $\chi$ leads to a second
wave front (denoted as mode D) with decreasing speed, which eventually turns non-ballistic at larger
$\chi$.  Since the wave front is eventually lost, we use the position of the peak of the density
profile $P^{\downarrow}_{x}$ to define a speed:
\[
v_{\text{peak}}(t) = \frac{1}{t} \left( \frac L2 - \underset{x \in [1, L/2]}{\mbox{argmax }} ( P^{\downarrow}_{x}(t) ) \right) \,,
\]
This is plotted in Fig.~\ref{fig:WFT}(right).  Up to $\chi \sim 0.5$, there is a linear decrease of
the speed, $v(J_y) \sim J_x - J_y$.  Around $\chi \sim 0.5$, the density profile $P^\downarrow_i(t)$
no longer shows an expanding two-peak structure but rather takes the form of a plateau
(Fig.~\ref{fig:WFT}, left), i.\,e. the ballistic wave front is lost.  Correspondingly,
$v_{\text{peak}}$ is seen to drop sharply to zero.  This is not due to an abrupt slowing down of
wave fronts, but rather due to the wave fronts becoming abruptly ill-defined.

Although the rung-specific mode does not propagate as wave fronts for $\chi \gtrsim 0.5$, there is
still spreading of the magnetization.  We characterize this process through the time-dependence of
the width of the plateau.
We consider the spatial standard deviation $\sigma$ of the normalized
density profile $P^\downarrow_i(t)/ P_0(t)$, given by
\begin{align} \label{eq:sigma_def}
    \sigma^2(t) = \frac {1}{P_0} \sum \limits_{i=1}^L (i - i_0)^2 \, P^{\downarrow}_i(t) \,,
\end{align}
with $P_0(t) = \sum_{i=1}^L P^\downarrow_i(t)$ and $i_0 = (1/P_0) \sum_{i=1}^L i\,
P^{\downarrow}_i(t) = L/2$, and fit it to $\sigma^2(t) = D t^{\alpha}$ (see App.~\ref{sup:TechRem} for further information).  An exponent $\alpha=2$ indicates
ballistic spreading, while $\alpha=1$ could be termed diffusive behavior. The fit gives sharp
results for the fitting coefficients $D$ and $\alpha$, see Fig~\ref{fig:Sigma}. For $\chi
\rightarrow \infty$, the coefficients approach values of the spin-1 chain, shown by the horizontal
lines in Fig.~\ref{fig:Sigma}.  The spreading of the signal gets slower with increasing $\chi$ until
around $\chi\approx1$ (both $\alpha$ and $D$ decrease).  For $\chi>1$ the exponent $\alpha$
increases rapidly to its asymptotic value, $\alpha(\chi\rightarrow\infty)\approx1.4$, while the
coefficient $D$ approaches its asymptotic value more slowly.  It is interesting that the spreading
is given by a well-defined power-law, which nevertheless is neither ballistic nor diffusive but
intermediate (``super-diffusive'').  At present, a detailed explanation of this anomalous diffusion
exponent is unavailable, but the spectral decomposition (next subsection) gives us some physical
understanding of the dynamics.

\subsection{The rung-specific mode: spectral decomposition} \label{sec:Rung_Spectral}
The coupling strength dependence of the rung-specific mode can be interpreted through the spectral
decomposition of the rung initial state (Fig.\ \ref{fig:MainRung} bottom row).  A remarkable feature
of this spectral decompostion is a branch \emph{within} the continuum.  This ``resonance'' branch is
responsible for the rung-specific dynamical mode.  (There is also spectral weight in the bound-state
branch of the $0s$ sector, corresponding to the weakly visible bound-triplet mode, and throughout the
bow-tie-shaped continua of both $0s$ and $2s$ sectors, corresponding to the single-particle
propagation mode.  We will not discuss these further.)

At small $\chi$, the branch within the continuum is very well-defined, as it approaches the form of the stable
single particle mode with a simple cosine dispersion at $\chi=0$. As a result, ballistic
behavior analogous to bound-pair propagation (e.g., the bound triplet mode or the two-magnon mode)
can be expected. This explains the apparently ballistic behavior we have observed at small $\chi$.  However,
since this branch is part of the continuum and thus has a finite width, i.\,e. lifetime, the propagating wave front 
decays exponentially (with time or with distance covered).  This is in contrast to the case of  particle
propagation associated with an energetically separated spectral branch with a $\delta$-function in energy, in 
which case the propagating wave front decays algebraically.  
%\textcolor{red}{[I believe this algebraic/exponential has not been checked
%    quantitatively numerically. Andreas: check if you are comfortable with the algebraic/exponential
%    assertion.]}
%    {\color{Blue} \textit{A quantitative check is found in the Dropbox: ``SpinLines/BiquadBilin/BqBl\_RspaceDecay.pdf''.}}
The ballistic mode at small $\chi$ should be regarded as the propagation of a
``resonance'' with a finite lifetime rather than the propagation of a stable particle or bound state mode.

Although the resonance branch is not a single sharp line, we can still loosely think in terms of a
``dispersion'' $\epsilon_k$.  As in bound pair propagation modes, the speed of ballistic propagation
for $\chi\lesssim0.5$ corresponds to the speed at the inflection point of this dispersion.  At
larger $\chi$, the branch broadens, i.\,e. the rung excitation hybridizes more strongly with the
continuum and there is more rapid decay of the resonance.  This coincides with the ballistic mode
disappearing and the signal width $\sigma(t)$ developing a non-ballistic exponent $\alpha<2$.  From
Fig.\ \ref{fig:MainRung} we see that the broadening happens first at the edges of the Brillouin zone
($k\sim 0, 2\pi$), and progressively extends to the center ($k\sim\pi$) with increasing $\chi$.  The
width of the branch around the inflection point presumably determines the time scale at which the
wave front decays.  For $\chi\gtrsim0.5$, the wave front does not survive up to the time scales ($t
\sim 100/J_x$) that we have used to determine the propagation speed.

As $\chi$ is increased, in addition to the increasing width, the curvature of the resonance branch
changes as well.  At $\chi \sim 1$ the branch dispersion has very little curvature and is almost
flat; hence dynamics is very slow.  This is roughly the regime where the spreading parameters
$\alpha$ and $D$ are minimal (Fig.\ \ref{fig:Sigma}), i.\,e. the spreading is slowest.  

For yet larger $\chi$, the dispersion shape actually gets inverted, and then gains slope in the
opposite direction.  This corresponds to increased $\alpha$ and $D$, i.\,e. faster spreading of
$\sigma(t)$, as is also visible in the rightmost panel in the center row of
Fig.\ \ref{fig:MainRung}.  We will analyze this strong-coupling regime further through current-like
operators in the next Subsection.

\subsection{Current operators and jamming dynamics}\label{sec:current}

%%%%%% FIGURE %%%%%% FIGURE %%%%%% FIGURE %%%%%% FIGURE %%%%%% FIGURE %%%%%%%%%%
\begin{figure}
  \centering
  \includegraphics[width=0.45\textwidth]{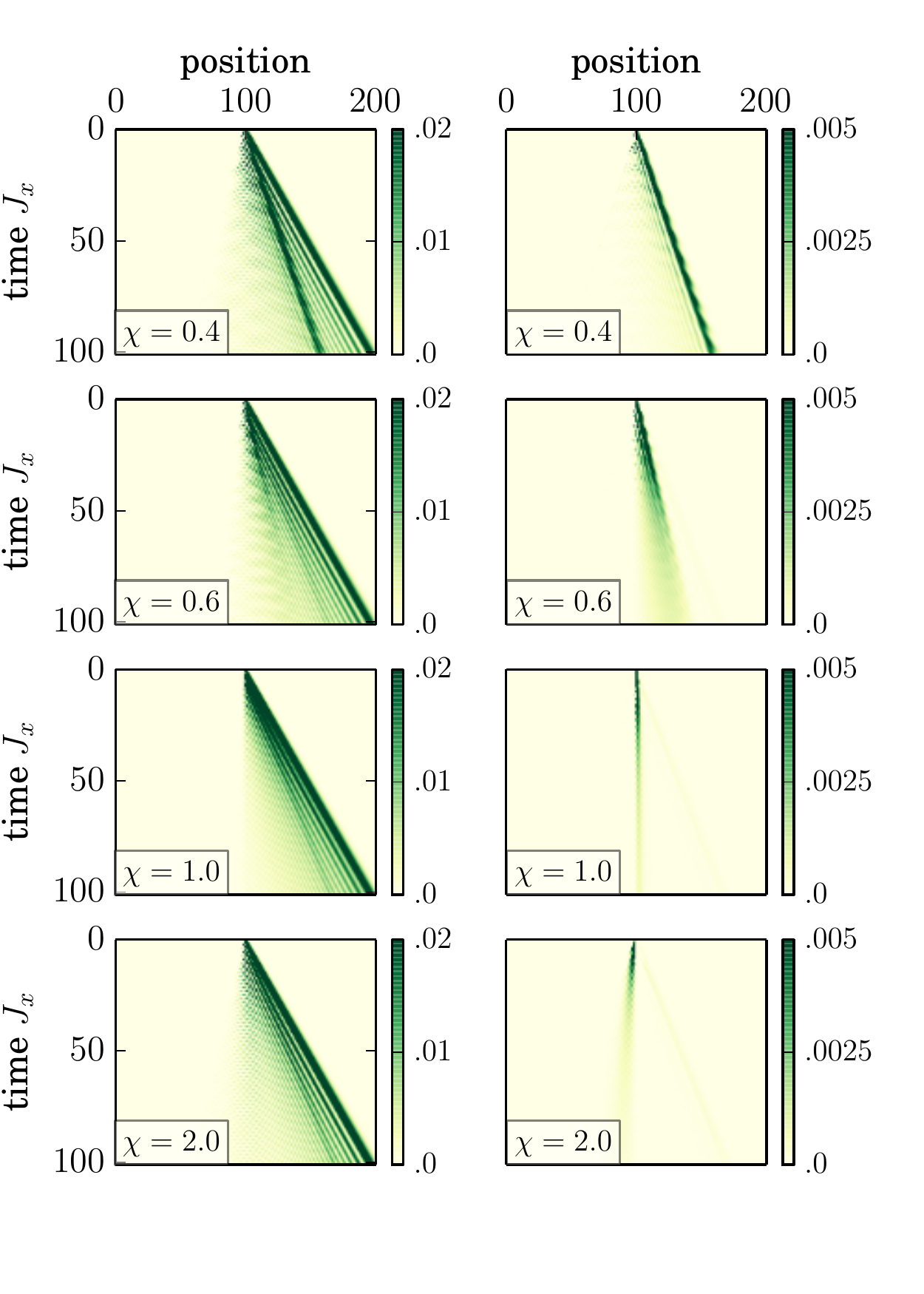}%CurrentLett+Phases.pdf}
  \caption{Current dynamics for the rung initial state.  Positive part of the current $j^{(1)}_i(t)$
    (first column) and positive part of the rung current $j^{(2)}_i(t)$ (second column) for
    different $\chi$ values.
  }
    \label{fig:Curr}
\end{figure}
%%%%%% FIGURE %%%%%% FIGURE %%%%%% FIGURE %%%%%% FIGURE %%%%%% FIGURE %%%%%%%%%%

Having described the magnetization dynamics of the rung-specific mode (\ref{sec:RungAnalysis}) and
the corresponding spectral picture (\ref{sec:Rung_Spectral}), we now analyze the dynamics in terms
of currents.  We regard the time evolution of the single particle current
\begin{align}
    j^{(1)}_{i}(t) = \mbox{Im } \langle S^+_{i,1} S^-_{i+1,1} + S^+_{i,2} S^-_{i+1,2} \rangle 
    \label{eq:Current1}
\end{align}
and the four-point Green's function
\begin{align}
    j^{(2)}_{i}(t) = \mbox{Im } \langle S^+_{i,1} S^+_{i,2} S^-_{i+1,1} S^-_{i+1,2} \rangle \, . 
    \label{eq:Current2}
\end{align}
The second operator represents physically the flow of $\ket{t_-}$ rungs.  It does not obey a
continuity equation like $\partial_{t}P^\downarrow_i(t) = \mbox{div }j_i$, because the number of
$\ket{t_-}$'s is not conserved in the dynamics; hence one has to be careful in interpreting this
quantity as a current.  (App.~\ref{sup:Current} discusses currents further.)  Note that the
subscript $i$ is used for the rightward current across the bond from $i$ to $i+1$. 

We compare both observables for the rung initial state in Fig.~\ref{fig:Curr}.  The current
functions are spatially antisymmetric, therefore, we put negative values to zero and show only
positive parts, for better visibility.  Thus, only rightward movement of magnetization and
$\ket{t_-}$ states is tracked in these Figures.  

Overall, the propagation of rung current $j_i^{(2)}$ resembles qualitatively the behavior of the
magnetization signal in the rung-specific mode; we see a slowing down as $\chi$ increases up to
$\chi\sim1$.  As $\chi$ is increased further, a peculiar effect appears: \emph{rightward} movement
is seen only on the \emph{left} half of the ladder.  In other words, occupied rungs ($\ket{t_-}$
states) seem to expand leftwards by moving rightwards, which would not be possible for the dynamics
of a conserved particle.  The resonace nature of this excitation, and the inversion of the
dispersion of the resonance branch in the strong-$\chi$ regime (discussed in
\ref{sec:Rung_Spectral}), lead to this unintuitive behavior.

%%%%%% FIGURE %%%%%% FIGURE %%%%%% FIGURE %%%%%% FIGURE %%%%%% FIGURE %%%%%%%%%%
\begin{figure}
  \centering
  \includegraphics[width=0.5\textwidth]{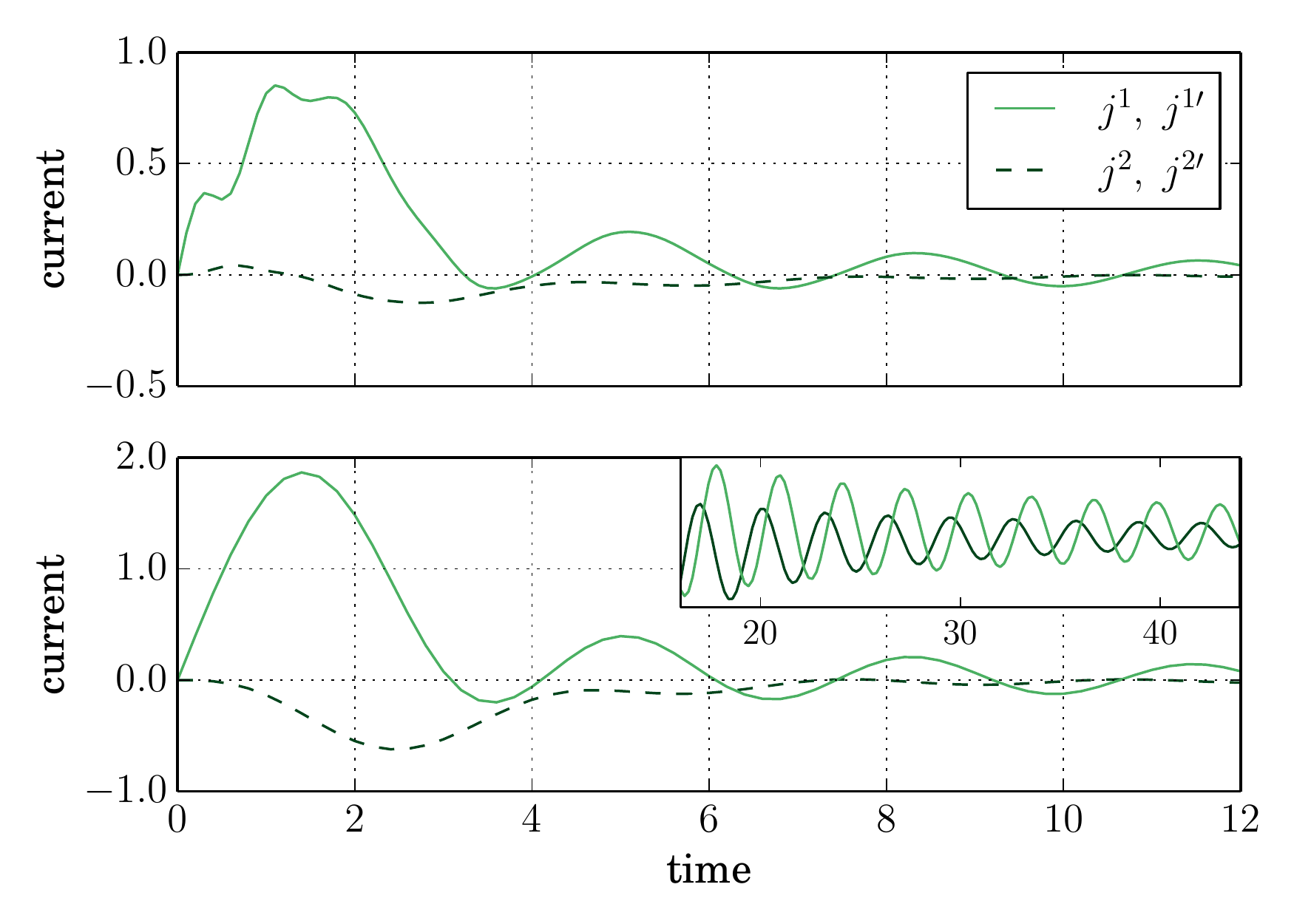}
  \caption{Comparison of current functions for the rung initial state at a strong coupling
    strength $\chi=4$ (top) and the corresponding situation of a spin-1 chain (coupling $J$),
    initialized as a single $S^z=-1$ site in an $S^z=1$ background (bottom) for systems with $L=101$
    rungs or lattice sites respectively.  Solid lines depict the current $j^{(1)}_{51} - j^{(1)}_{50}$
    (top) and $j^{(1')}_{51} - j^{(1')}_{50}$ (bottom) and dashed lines show the rung (or spin-1)
    current $j^{(2)}_{51} - j^{(2)}_{50}$ (top) and $j^{(2')}_{51} - j^{(2')}_{50}$ (bottom) from the
    position where both initial states differ from the ferromagnetic background.  The inset displays
    the shifted oscillations of both observables at later times (for a better comparison in the
    inset, $j^{2'}$ has been normalized with a factor of eight).  Time is given by $t / J_x$ for the
    top and $t 2/ J_{\mbt{bl}}$ for the bottom Figure (App.\ \ref{sup:BBMmap}).  }
    \label{fig:Jamming}
\end{figure}
%%%%%% FIGURE %%%%%% FIGURE %%%%%% FIGURE %%%%%% FIGURE %%%%%% FIGURE %%%%%%%%%%

In Fig.\ \ref{fig:Jamming}(top panel), we show $j^{(1)}_{\ell_0}-j^{(1)}_{(\ell_0-1)}$, where $\ell_0$ is
the index of the central rung where the two upturned spins are initially placed.  This is the
outgoing current from the central rung.  We compare with $j^{(2)}_{\ell_0}-j^{(2)}_{(\ell_0-1)}$, the
outgoing current of $\ket{t_-}$ states.

In the lower panel, we compare analogous quantities in the spin-1 BLBQ chain (at the Heisenberg
point $J_{\mboti{bq}}=0$).  The currents are $j^{(1')}_{i} =\mbox{Im }\langle S^+_i S^-_{i+1}\rangle$
and $j^{(2')}_{i} =\mbox{Im }\langle S^+_{i}S^+_{i} S^-_{i+1} S^-_{i+1}\rangle$, the initial state is
a single $S^z=-1$ site in an $S^z=1$ background.  We display $j^{(1')}_{\ell_0} -
j^{(1')}_{(\ell_0-1)}$ and $j^{(2')}_{\ell_0}-j^{(2')}_{(\ell_0-1)}$, indicating outgoing magnetization
current and outgoing current of double-occupancy from the central site $l_0$.  The behavior in upper
and lower panels are very similar, indicating that the spin-1 chain is an excellent effective model
for the rung dynamics in this large-$\chi$ regime.  We exploit this mapping in detail in the
following Section.

At short times, the positivity of the outflow of $j^{(1)}$ ($j^{(1')}$) implies single-particle
magnetization dynamics away from the central rung (site).  The $\ket{t_-}$ current (current of
double occupancy) has opposite (i.\,e. inward flowing) behavior, as seen by the negative values of
$j^{(2)}$ ($j^{(2')}$).  In the inset, we show longer time scales; the behavior is very similar for the
two models.  We notice that the two quantities oscillate with the same frequency, but not in phase:
there is a phase shift of approximately $\pi/2$.  This quantifies the jamming mechanism involving
the counter-propagation of two types of excitations.

\begin{figure}
  \centering
  \includegraphics[width=0.48\textwidth]{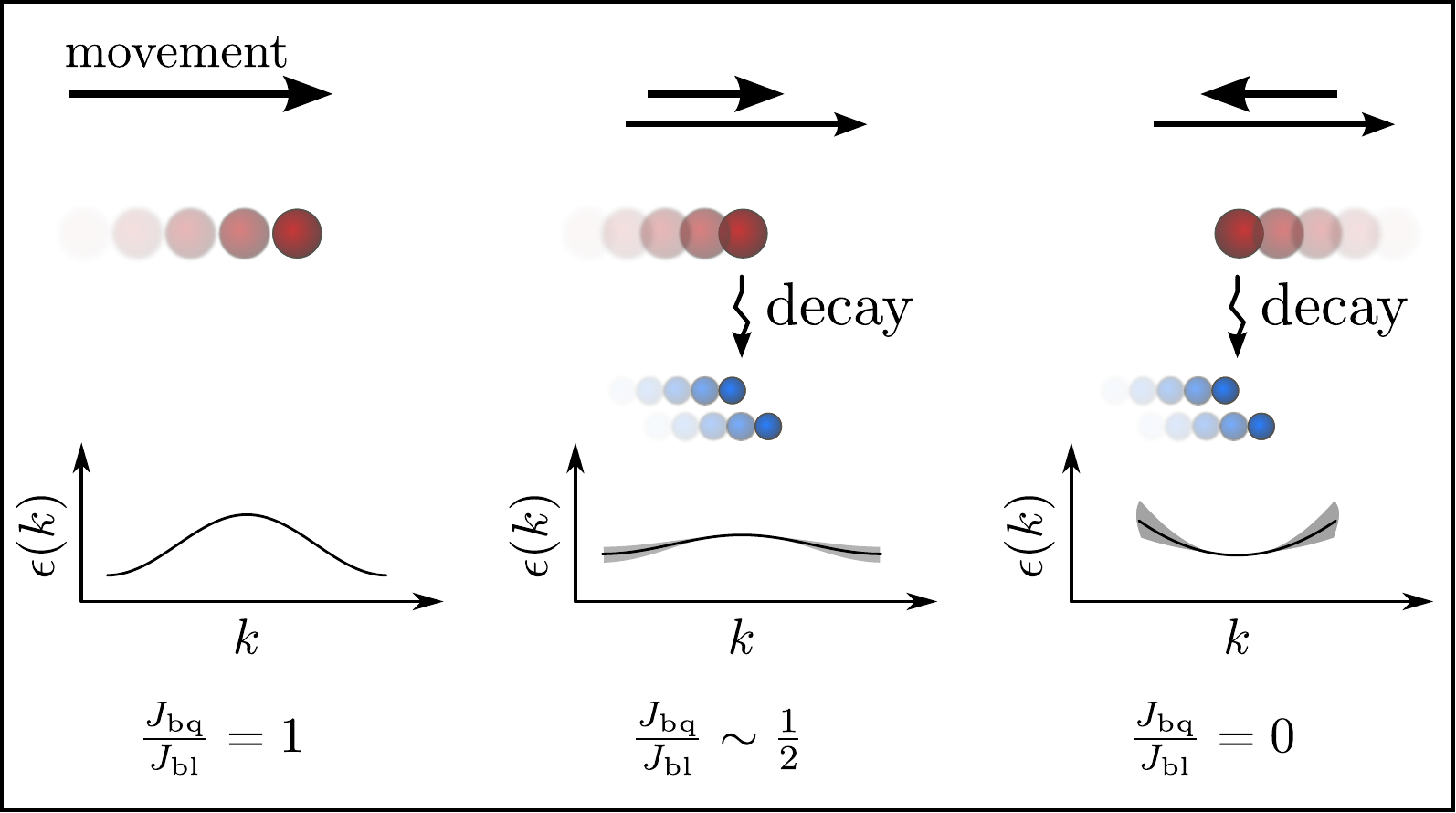}
  \caption{Heuristic picture of jamming dynamics for the BLBQ chain.  Movement of doubly occupied
    sites is denoted by red spheres and bold arrows while expansion of single magnon states is
    indicated by blue spheres and thin arrows.  Sketches of the dispersion $\epsilon(k)$, with the
    crucial features of reshaping and broadening, are at the bottom of the Figure.  The situations
    for three different couplings are displayed.  At the $SU(3)$ symmetric point, $J_{\mbt{bq}}/
    J_{\mbt{bl}} =1$, the quasiparticle corresponding to double occupation ($S^z=-1$ sites) has a
    sharp cosine dispersion.  Broadening of the dispersion for smaller $J_{\mbt{bq}}/ J_{\mbt{bl}}$
    enables decay of the quasiparticle to two-magnon scattering states, $J_{\mbt{bq}}/ J_{\mbt{bl}}
    \sim 0.5$.  In addition to the emergence of decay, the expansion velocity of the quasiparticle,
    as given by the curvature of the dispersion, also changes.  Near the Heisenberg point
    $J_{\mbt{bq}}/ J_{\mbt{bl}} \sim 0$, the quasiparticle dispersion has inverted its curvature
    hence reversing the direction of expansion.  This results in the counterpropagation of the
    quasiparticle and its decay products.  }
    \label{fig:JammingComic}
\end{figure}

Results for the spectral changes of the rung initial state (Sec.~\ref{sec:Rung_Spectral}) and for
the behavior of current functions discussed above may be combined into a simple heuristic picture to
explain the peculiar dynamical behavior of the rung-specific mode as a function of $\chi$.  This
picture is summarized in Fig.~\ref{fig:JammingComic}.  It is simpler to use the spin-1
bilinear-biquadratic (BLBQ) chain (App.~\ref{sup:BBMmap}) to describe the jamming mechanism,
compared to the more complicated spin-$\frac 12$ ladder.  As seen above, the BLBQ chain shows
similar dynamics as the rung-specific mode when initiated with a single $S^z=-1$ site in a
ferromagnetic $S^z=1$ background.  
For $J_{\mboti{bq}} = 0$ (Heisenberg point), the BLBQ chain corresponds to the $\chi\gg1$ limit of
the ladder.  At $J_\mboti{bq} = J_\mboti{bl}$, the BLBQ is the $SU(3)$ symmetric \textit{permutation
  model} (App.~\ref{sup:BBMmap}), which has similar behavior to the small-$\chi$ behavior of the
ladder.  Increasing the rung coupling $\chi$ in the ladder system is thus analogous to decreasing
$J_\mboti{bq}$ from $J_\mboti{bq}=J_\mboti{bl}$ to $J_\mboti{bq}=0$ in the BLBQ.
Fig.~\ref{fig:LadBBMCompRung} compares the spectral decompositions of the initial state for
different $J_\mboti{bq}$ in the BLBQ chain and different $\chi$ in the ladder.  Comparison of
real-space behavior shows a similar correspondence of parameters in the two models.

Close to the $SU(3)$ point $J_{\mbt{bq}} /J_{\mbt{bl}} \sim 1$, the spectral decomposition shows a
narrow dispersion.  The dynamics involves coherent propagation of quasiparticles, with ballistic
wave fronts having velocity given by Eq.~(\ref{eq:DosArg}).  Considering the two current functions
analyzed above, the expansion of these quasiparticles is given by propagating $S^z=-1$ sites (or
doubly occupied sites in the particle language).  The key feature leading to jamming dynamics is the
fact that this dispersion is realized with overlap of states from the scattering continuum.
Decreasing the biquadratic coupling results in two effects: First, a change of curvature of the
disperion, associated with a change of expansion velocity.  Second, a broadening of the dispersion
such that interference of many energetically close scattering states leads to enhanced spatial decay
of expanding wave fronts.  A dispersive broadening is connected to decay of the formerly (at the
$SU(3)$ point) well-defined quasiparticles.  So the ``doublon'' quasiparticle associated with the
double occupation of a lattice site decays into magnon scattering states during time evolution.  In
contrast to the doublon, expansion of these scattering states (as diplayed by the current functions
$j^{(1)} (j^{(1')})$, Fig.~\ref{fig:Curr}, left column) do not change qualitatively when the
coupling is changed.  Other than the broadening of the quasiparticle dispersion, its curvature
changes sign for low enough biquadratic coupling, leading to an inversion of expansion velocity
according to Eq.~(\ref{eq:DosArg}).  The jamming phenomenon is thus caused by the counterpropagation
of the quasiparticle corresponding to doubly occupied sites and its decay products.

%%%%%% FIGURE %%%%%% FIGURE %%%%%% FIGURE %%%%%% FIGURE %%%%%% FIGURE %%%%%%%%%%
%%%%%% FIGURE %%%%%% FIGURE %%%%%% FIGURE %%%%%% FIGURE %%%%%% FIGURE %%%%%%%%%%

We next present an analytical treatment, via perturbation theory, of the two spectral effects
associated with the rung-specific mode as a function of $\chi$: namely the broadening and the change
of curvature of the ``resonance'' branch.

% We regard the spectral decomposition of the rung initial state for $J_y/J_x > 1$: Considering the
% $S^z$ basis and its division into rung, leg and scattering states we note that dynamics of the
% rungs is given by a second order process with leg states as intermediate steps.  From both current
% functions we learn, that scattering states prefer a movement into one direction, while rung states
% exclusively move to the opposite direction.  This is reflected in the curvature of the dispersion
% obtained from the spectral decomposition.  As a consequence, a rung initial state is confined to
% its initial position since the scattering states provoke a jamming by counteracting dynamics.
% Beyond this heuristic interpretation, we have no explanation for this peculiar observation.

\subsection{Perturbative treatment of spectral features using correspondence to BLBQ chain} \label{sec:BLBQ} 
\begin{figure}
  \centering \includegraphics[width=0.5\textwidth]{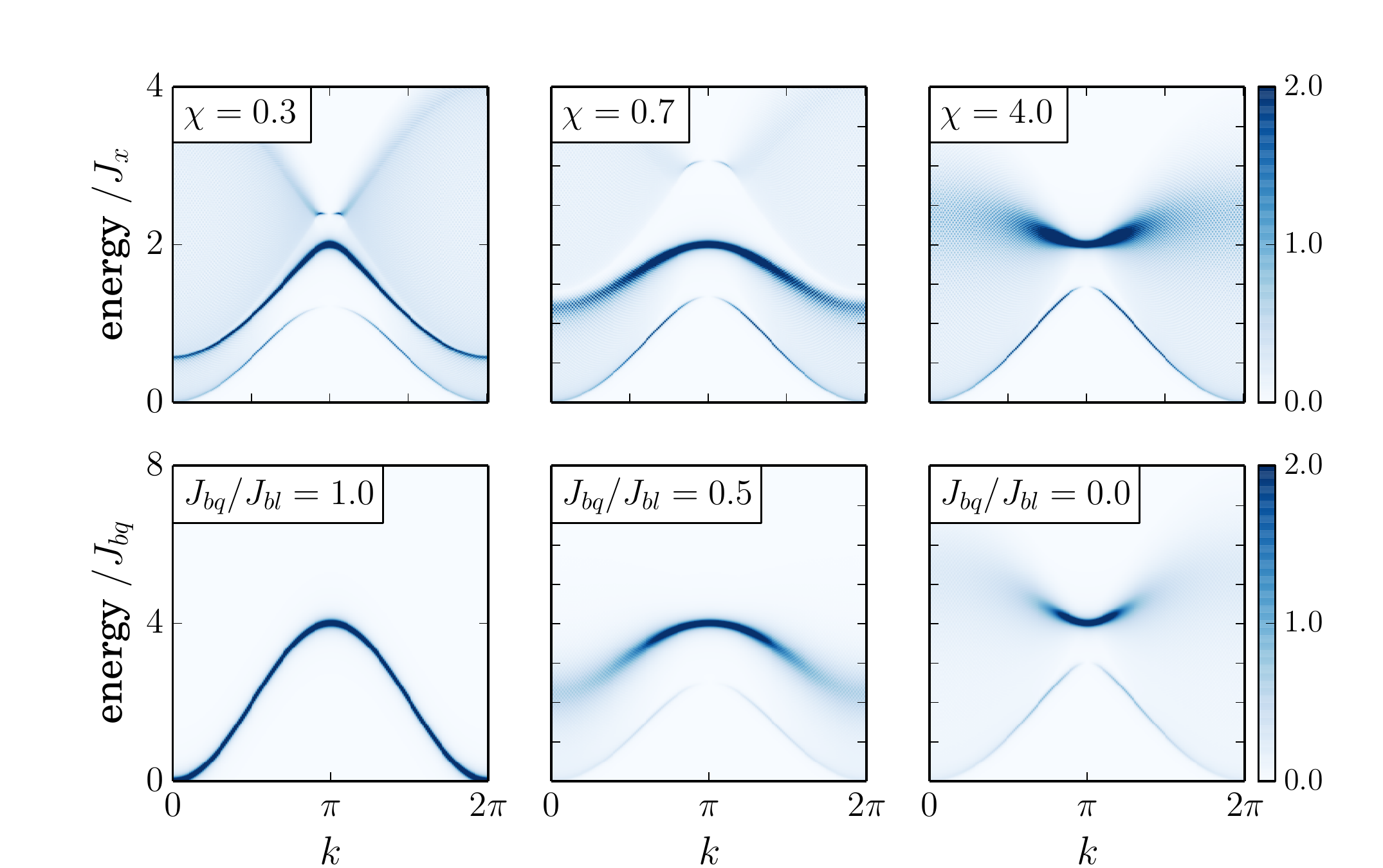}
  \caption{Comparison of the spectral decomposition of the rung initial state in the $0s$ sector
    (first line) and the corresponding initial state in the BLBQ (second line).  In both cases a
    well-defined dispersion reduces its curvature by lifting at the edges and broadens to enable
    transitions to a scattering subspace.  }
    \label{fig:LadBBMCompRung}
\end{figure}

We have seen two nontrivial effects in the dynamics of the rung intial state: the loss of wave
fronts at coupling strength $\chi \sim 0.5$ and the non-monotonic behavior of expansion velocity as
a function of $\chi$ as quantified through the behavior of $\sigma^2(t)\sim Dt^{\alpha}$.  The
corresponding spectral decomposition of the rung initial state shows that the first phenomenon is
connected to the broadening of the dispersion, while the second phenomenon is associated with the
change of curvature of the dispersion.  We now provide an analytical description of both effects
using perturbation theory.  Again, we work with the simpler BLBQ chain.
The two spectral effects are seen also in the BLBQ chain and are very similar to the effects in the
spin-$\frac 12$ ladder, Fig.~\ref{fig:LadBBMCompRung}.  Starting from the $SU(3)$ point (permutation
model), we will examine these two effects perturbatively.
We summarize the results here; calculation details are in App.~\ref{sup:Fermi}.

%%%%%% FIGURE %%%%%% FIGURE %%%%%% FIGURE %%%%%% FIGURE %%%%%% FIGURE %%%%%%%%%%
%%%%%% FIGURE %%%%%% FIGURE %%%%%% FIGURE %%%%%% FIGURE %%%%%% FIGURE %%%%%%%%%%

The Hilbert space of the permutation model with two excitations is composed of two separated
subspaces.  One subspace is composed of configurations with a single $S^z=-1$ site in a
ferromagnetic $S^z=1$ background (both excitations on the same site).  This space is trivially
mapped to the Hilbert space of a single particle.  The other, disjoint subspace contains
configurations where the two excitations are on different sites; this subspace maps to the
spin-$\frac 12$ Heisenberg chain.  The separation of these two subspaces makes the $SU(3)$ point
integrable and relatively simple.  The initial state at this point is in the first (single-particle)
subspace.  This explains the purely cosine-shaped sharp spectral decomposition at this point
(Fig.~\ref{fig:LadBBMCompRung} bottom leftmost panel).

\begin{figure}
  \centering
  \includegraphics[width=0.5\textwidth]{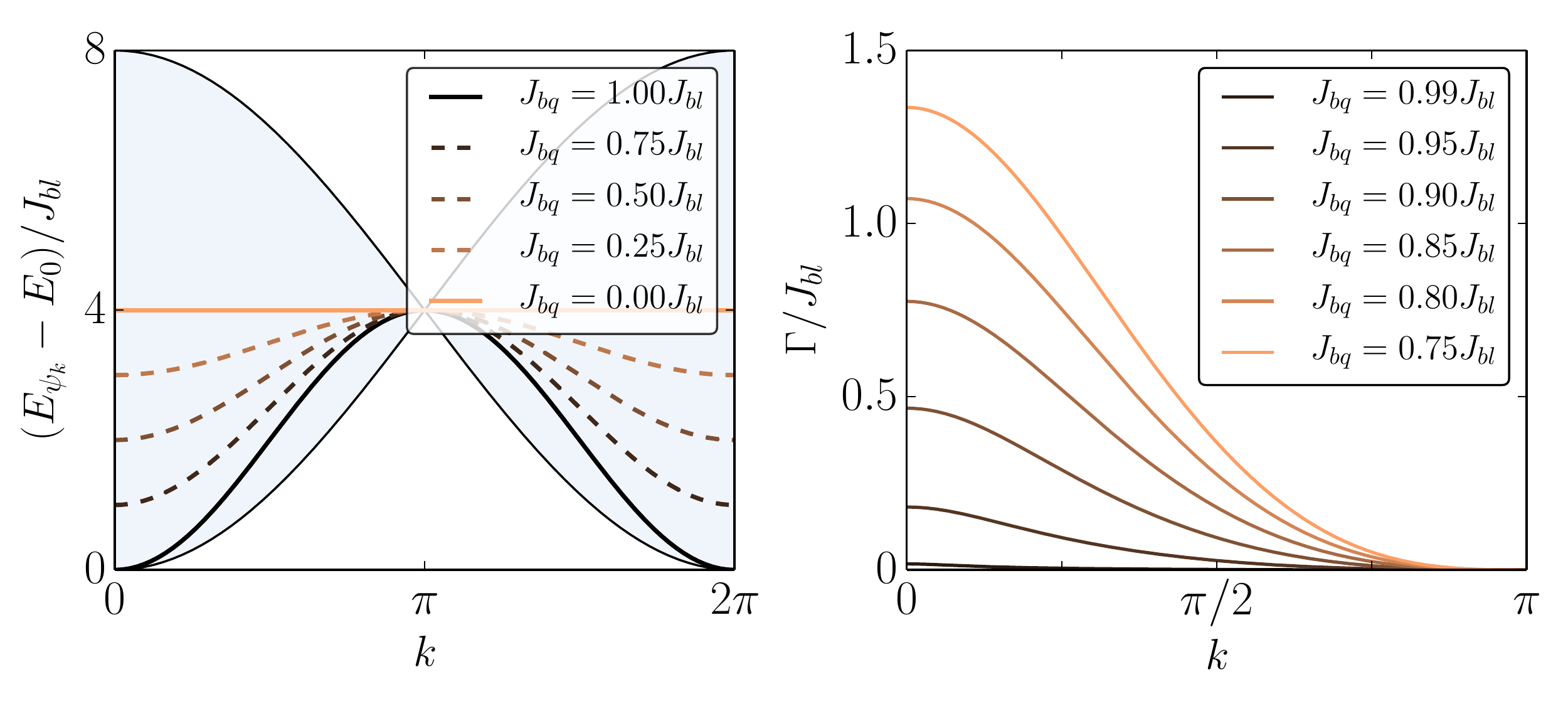}
  \caption{ Perturbative results for the BLBQ chain, for the two spectral effects associated with
    the rung-specific mode.  Left: Energy (dispersion) for different $J_\mboti{bq}$.  The dispersion
    starts as a cosine at the $SU(3)$ point and lifts at the edges of the Brillouin zone with
    decreasing $J_\mboti{bq}$.  At first order we do not obtain a true inversion of the dispersion,
    but a flat dispersion is obtained at the Heisenberg point.  Right: Decay width.  The decay
    increases with $J_\mboti{bq}$.  States around $k \approx \pi$ are stable and instability
    increases for states with lower group velocities.  }
    \label{fig:SemiAnalytics}
\end{figure}

When tuning the BLBQ from the $SU(3)$ point towards the Heisenberg point, there is a mixing of
states from the single particle sector with states from the rest of the Hilbert space.  
By calculating the first order correction to the energy of our initial state, we can explain the
change of curvature of the dispersion.    The energy expectation value of the
Fourier transformed components of this state $\ket{\psi_k}$ are given by
\begin{align}
    E_{\psi_k} - E_0 = 4 J_\mboti{bl} - J_\mboti{bq} ( 2 + 2\, \mbox{cos }k )
    \label{eq:rung_energy}
\end{align}
and the ground state energy $E_0 = -L (J_\mboti{bl} + J_\mboti{bq})$.  A decrease of $J_\mboti{bq}$
leads to a lifting of the energy, hence to the change of curvature of the dispersion.  A plot of
this first-order-corrected energy dispersion is shown in Fig.~\ref{fig:SemiAnalytics}.  In the
perturbative analysis, the dispersion becomes flat at the Heisenberg point $J_\mboti{bq}=0$; the
actual resonance branch assumes a flat shape already around $J_\mboti{bq}/J_\mboti{bl} \approx 0.3$
(Fig.~\ref{fig:LadBBMCompRung} bottom row).

The broadening of the resonance branch can also be calculated using perturbation theory around the
$SU(3)$ point.  Since the broadening is due to coupling to the continuum, at first order it is given
by Fermi's Golden Rule (App.~\ref{sup:Fermi}).  Results are shown in
Fig.~\ref{fig:SemiAnalytics}(right panel).  The broadening is large near the edges of the Brillouin
zone ($k=0$ and $k=2\pi$) and small near the center ($k=\pi$), as also observed in the full
numerical calculations (Fig.~\ref{fig:MainRung}).

While the uncoupled ($\chi=0$) ladder and the BLBQ at the $SU(3)$ point are not equivalent, changes
in the relevant parameter (increasing $J_y$ or decreasing $J_\mboti{bq}$) provoke a crossover from
ballistic to non-ballistic propagation dynamics in both cases.  The simple structure of the $SU(3)$
point allows the perturbative calculation presented here.  For the BLBQ chain, one can interpret the
broadening phenomenon as a breaking of integrability: at the integrable $SU(3)$ point, different
excitation branches are decoupled so that the $S_z=-1$ branch has a sharp identity.  Away from this
point, different types of excitations get hybridized, breaking the integrability and also
hybridizing the branch with the continuum.

%%%%%% FIGURE %%%%%% FIGURE %%%%%% FIGURE %%%%%% FIGURE %%%%%% FIGURE %%%%%%%%%%
%%%%%% FIGURE %%%%%% FIGURE %%%%%% FIGURE %%%%%% FIGURE %%%%%% FIGURE %%%%%%%%%%

\section{Conclusion}\label{sec:Conc}

In this work we have studied the modes of propagation in a spin-$\frac 12$ Heisenberg ladder, generated
from two overturned, adjacent spins in an otherwise polarized ladder.  The case of two excitations
already provides a variety of effects, with four fundamentally different modes of propagation.
Depending on the coupling strength regime, subspaces of the Heisenberg ladder is mapped to different
simpler models.  We have exploited these mappings to interpret the nature of the different modes.
Concidering the spectral decompositions of the initial states into the eigenstates of the system, we
have achieved a classification of the various modes of propagation in terms of different propagating
quasiparticles.  

Other than a single-particle mode which is present with all initial states, the other three modes
are all nontrivial collective effects. 
For the leg initial state, we observed ballistic propagation of two-string bound complexes (familiar
from the $XXZ$ chain \cite{Ganahl2012a,Vlijm2015}) as well as propagating triplet $\ket{t_0}$ bound
states.  The rung initial state shows the propagation of a resonance which ceases to propagate
ballistically for coupling strengths $\chi \gtrsim 0.5$.  In the large $\chi$ regime, through
analysis of different types of currents we have found that this peculiar `jamming' effect is due
counterpropagation of two types of quasiparticles.  The peculiar dynamics is reflected in nontrivial
changes of a resonance branch in the spectrum as a function of $\chi$.  We have found the same
phenomenon in the spin-1 bilinear biquadratic (BLBQ) chain, to which the ladder can be mapped onto
in the large  $\chi$ regime. Exploiting this correspondence, we have used perturbative calculations
on the BLBQ chain to explain the spectral effects.

It is expected that unitary propagation phenomena in two-dimensional quantum lattices will show a
far richer zoo of nontrivial effects, compared to the phenomena now known for the better-studied
one-dimensional case.  The present study can be seen as a first step in this direction.  Indeed,
just with a ladder rather than a full 2D structure, and confining ourselves to just the two-particle
sector, has already led us to the rich collection of phenomena we have reported in this paper.  In
particular, the jamming phenomenon associated with the rung initial state is qualitatively different
from anything we are aware of in the literature.  The present work is thus expected to lead to many
new phenomena in the field of propagation, jamming and interactions between propagating modes in
geometries beyond the simplest chain geometries.

\acknowledgments

We thank I.~Bloch, C.~Gross and A.~Rosch for stimulating discussions. CK and AML acknowledge support 
by the Austrian Science Foundation FWF through the SFB FoQus (F4018).

% \bibliography{SpinDynPaper}

% - - - - - - - - - - -
%%% A P P E N D I X %%%
% - - - - - - - - - - -
\appendix

\section{Mappings to sub-Hilbert spaces}

In this work we have employed a number of mappings of the ladder to simpler systems (in particluar
to chains) to elucidate aspects of the dynamics.
In the follwing subsections, we outline these mappings: the mapping of the $0s$ subspace at large
$\chi$ to the spin-1 BLBQ chain (\ref{sup:BBMmap}), the mapping of the $1s$ subspace to an
anisotropic ($XX$) and an isotropic (Heisenberg or $XXX$) spin-$\frac 12$ chain (\ref{sup:AntiMap}), the
mapping of the $2s$ subspace (also at large $\chi$) to the anisotropic ($XXZ$) spin-$\frac 12$ chain
(\ref{sup:MapD12}), and finally the mapping of the $\chi=0$ case to a non-interacting chain
(\ref{sup:LowChiMap}).

\subsection{Mappings of the $0s$ subspace} \label{sup:BBMmap}

We describe below how, for strong coupling $\chi\gg1$, the dynamics of the $0s$ subspace of the
spin-$\frac 12$ ladder turns into that of the spin-1 BLBQ chain.  
%
%% The physics of the spin-1 chain is reflected in the bound triplet mode for the leg initial state and
%% also in the peculiar dynamics of the mode specific to the rung initial state.

A perturbative approximation starting from the large $\chi$ limit, used to project onto the triplet
subspace of the ladder, yields the spin-1 Heisenberg chain at lowest order (as already known in the
literature \cite{Mila1998}) and the biquadratic term at the next order, so that one obtains the BLBQ
chain as a good approximation down to moderate values of $\chi$.

We divide the Hilbert space into states built as a product of triplets
\begin{align*}
    \mathcal{P} = \{ \ket{t_-}, \ket{t_0}, \ket{t_+} \}
\end{align*}
and states containing at least one singlet $\ket{s}$ rung.  We will project the Hamiltonian onto the
first subspace (``model space'').   The resulting Hamiltonian can be interpreted as a spin-1 chain
with the natural identification befween rung states of the ladder and site states of the chain:
\[
     \ket{t_+} \leftrightarrow \ket{S^z=1},\quad
     \ket{t_0} \leftrightarrow \ket{S^z=0},\quad
     \ket{t_-} \leftrightarrow \ket{S^z=-1}, 
\]
so that the spin-1 operator ${\bf T}$ acting on a rung will have the following actions: 
\begin{align*}
    T^+ \ket{t_m} &= (1/\sqrt{2})\ \ket{t_{m+1}}, \quad T^+ \ket{t_{+}} = 0, \\
    T^- \ket{t_m} &= (1/\sqrt{2})\ \ket{t_{m-1}}, \quad T^- \ket{t_{-}} = 0, \\
    T^z \ket{t_m} &= m \ket{t_m} .
\end{align*}

The Hamiltonian \eqref{eq:Hamiltonian} has a rung part (with couplings $J_y$) and a leg part (with
couplings $J_x$):
\begin{align}
    H = H_{\mbt{rung}} ~+~ H_{\mbt{leg}} \,.
    \label{eq:DecoH}
\end{align}
%% \begin{align}
%%     H = \underbrace{-J_y \sum \limits_{i=1}^L {\bf S}_{i,1} \cdot {\bf S}_{i,2} }_{H_{\mbt{rung}}}\ \underbrace{-J_x \sum \limits_{y=1}^2 \sum \limits_{i=1}^{L} {\bf S}_{i,y} \cdot {\bf S}_{i+1, y} }_{H_{\mbt{leg}}} \,.
%%     \label{eq:DecoH}
%% \end{align}
We treat   $H_{\mbt{rung}} = H_0$ as the unperturbed part and $H_{\mbt{leg}} = H_1$ as the perturbation.
In the unperturbed limit $(H_1=0)$, the system is a collection of decoupled rungs.
All ladder states built from rungs of the set $\mathcal{P}$ are eigenstates of $H_0$ and
degenerate. 
% $H_0 \ket{p} = \epsilon \ket{p}, \forall \ket{p} \in \mathcal{P}$.

To perform the perturbative projection, we define $P$ to be the projector onto the model space and
$Q=1-P$, with $P^2=P$ and $Q^2=Q$.  the full Schr{\"o}dinger equation $H \ket{\psi_i} = E_i
\ket{\psi_i}$ is decomposed as \cite{Fulde1991} 
\begin{align}
    &\left| \begin{array}{r} PHP \ket{\psi_i} + PHQ \ket{\psi_i} = E_i P \ket{\psi_i} \\
        QHP \ket{\psi_i} + QHQ \ket{\psi_i} = E_i Q \ket{\psi_i} \end{array} \right| \nonumber \\
        \Leftrightarrow 
        &\left| \begin{array}{r} [PHP + PHQ (E_i - QHQ)^{-1} QHP] \ket{\psi_i} = E_i P \ket{\psi_i} \label{eq:HeffSGL}\\
            (E_i - QHQ)^{-1} QHP \ket{\psi_i} = \phantom{E_i} Q \ket{\psi_i} \end{array} \right| \,.
\end{align}
The first line defines a Schr{\"o}dinger equation for the model space with an unaltered energy
spectrum $H_\mboti{eff} P \ket{\psi_i} = E_i P \ket{\psi_i}$ for the effective Hamiltonian
\begin{align*}
    H_\mboti{eff} = PHP + PHQ \frac{1}{E_i - QHQ} QHP \,.
\end{align*}
Expanding the fraction according to $\frac{1}{A-B} = \frac{1}{A} \sum
\limits_{n=0}^\infty \left( B\, \frac{1}{A} \right)^n$, with 
$A = \epsilon - QH_0Q$ and $B = QH_1Q -E_i +\epsilon$, gives
\begin{multline}    \label{eq:RSExpansion}
    H_\mboti{eff} = PHP \\
        + PH_1Q \frac{1}{\epsilon - H_0}\, \sum \limits_{n=0}^\infty \left( (QH_1 - E_i + \epsilon)\, \frac{1}{\epsilon - H_0} \right)^n QH_1P \,,
\end{multline}
which constitutes the Rayleigh Schr{\"o}dinger perturbation series.
Omitting the second term in the expansion  (\ref{eq:RSExpansion}) leads to the first order
approximation: 
\begin{align}
   H_{\mbt{eff}}^{(1)} =  PHP =   H_0 + P H_1 P \,.
    \label{eq:PerturbH}
\end{align}
Using this and translating to the  ${\bf T}$ operators we obtain, up to a constant, 
\begin{align*}
    H_{\mbox{\tiny{eff}}}^{(1)} = -\frac{J_\mboti{bl}}{2} \sum_{\langle i,j \rangle} \left\{T_i^+ T_j^- +
    \mbox{h.\,c.}\right\} - J_\mboti{bl} \sum_{<i,j>} T_i^z T_j^z
\end{align*}
with $J_\mboti{bl} = J_x/2$.  This is a spin-1 Heisenberg chain. 

The second order is obtained by including the first ($n=0$) term of the summation in
\eqref{eq:RSExpansion}.  This leads to a biquadratic term in the effective Hamlitonian. 
The matrix elements of $H_\mboti{eff}$ up to order $n=0$ ($\epsilon = -J_y/2$), are 
\begin{align*}
    &\bra{t_+t_+} H_\mboti{eff} \ket{t_+t_+} = -\frac{J_y}{2} -\frac{J_x}{2}, \\ &\bra{t_0t_0} H_\mboti{eff} \ket{t_0t_0} = -\frac{J_y}{2} - \frac{1}{8}\, \frac{J_x^2}{J_y}, \\ 
    &\bra{t_+t_0} H_\mboti{eff} \ket{t_+t_0} = -\frac{J_y}{2}, \\ &\bra{t_+t_-} H_\mboti{eff} \ket{t_+t_-} = -\frac{J_y}{2} + \frac{J_x}{2} - \frac{1}{8}\, \frac{J_x^2}{J_y},
\end{align*}
(diagonal terms) and
\begin{align*}
    &\bra{t_+t_-} H_\mboti{eff} \ket{t_0t_0} = -\frac{J_x}{2} + \frac{1}{8}\, \frac{J_x^2}{J_y}, \\ &\bra{t_0t_+} H_\mboti{eff} \ket{t_+t_0} = -\frac{J_x}{2}, \\ 
    &\bra{t_+t_-} H_\mboti{eff} \ket{t_-t_+} = - \frac{1}{8}\, \frac{J_x^2}{J_y} .
\end{align*}
(finite off diagonal terms).  The missing matrix elements are deduced by performing a $\mathbb{Z}_2$
operation ($t_+ \leftrightarrow t_-,\ t_0 \leftrightarrow t_0$), which leaves the Hamiltonian and
its matrix elements unchanged.  It is sufficient to consider two neighboring rungs; next-nearest
neighbor terms are not generated at this order.

%% It is sufficient to show the equivalence for a system with two rungs
%% as the Hamiltonian is a sum of single site and nearest neighbor operators and we expand to second
%% order.  Although the second term of $H_\mboti{eff}$ is a product of operators with $H_1$ showing up
%% twice, there are no finite matrix elements corresponding to next nearest neighbor hopping.  The
%% reason for this is that such a term has to be achieved by a virtual process leading out of the
%% unperturbed triplet subspace ($QH_1P$).  The second virtual hopping process ($PH_1Q$) has to involve
%% a transition back to the unperturbed subspace. Consider any state of $\mathcal{P}$ and the first
%% virtual process acting on site two and three, like $\ket{t\, t\, t} \rightarrow \ket{t\, s\, s}$.
%% %A combination of fully polarized triplets is the only case where a simultaneous change of total
%% spin and magnetization is realized by acting with $H_1$, meaning a process like $\ket{t\, t_{\pm}
%% t_{\mp}} \rightarrow \ket{t\, s\, s}$, where the first rung is an unassigned triplet. 
%% To have a finite next nearest neighbor term, the second process has to act on the first two sites.
%% To obtain a finite matrix element, the second process has to lead back to the unperturbed space
%% hence acting on site two and three.  Such a process is not possible, so we exclude three site terms
%% and continue with the relevant two site matrix elements displayed above.

Using these matrix elements, the effective Hamiltonian takes the form of the  spin-1 bilinear
biquadratic chain 
\begin{align*}
    H_\mboti{eff} = -J_\mboti{bl} \sum \limits_{i} {\bf T}_i \cdot {\bf T}_{i+1} - J_\mboti{bq} \sum
    \limits_{i} \left( {\bf T}_i \cdot {\bf T}_{i+1} \right)^2 
\end{align*}
with the couplings $J_\mboti{bl} = J_x/2$ and $J_\mboti{bq} = J_x^2/(8 J_y)$.  There is a constant
energy shift $(\frac 14 J_\mboti{bl}^2/ J_\mboti{bq} -J_\mboti{bq})L$, omitted above.

The BLBQ chain can also be expressed in terms of permutation operators: 
\begin{align}
H &= -J_\mboti{bl} \sum \limits_{\langle i,j\rangle} {\bf T}_i \cdot {\bf T}_j -
J_\mboti{bq} \sum \limits_{\langle i,j\rangle} \left( {\bf T}_i \cdot {\bf T}_j
\right)^2 \notag \\ 
&= -J_\mboti{bq} \sum \limits_{\langle i,j\rangle} \left( 1 + P_{i,j} \right) -
(J_\mboti{bl} - J_\mboti{bq}) \sum \limits_{\langle i,j\rangle} {\bf T}_i \cdot {\bf
  T}_j . 
\label{eq:HamiltonBBM}
\end{align}
Here we used the relation $P_{i,j} = ({\bf T}_i \cdot {\bf T}_j)^2 + {\bf T}_i \cdot {\bf T}_j - 1$
to express the Hamiltonian with the permutation operator exchanging the content of two sites
\begin{align*}
    P_{i,j} \ket{\alpha}_i \otimes \ket{\beta}_j = \ket{\beta}_i \otimes \ket{\alpha}_j
\end{align*}
with $\alpha, \beta \in \{-1, 0, 1\}$ denoting the spin component $S^z$ on the individual sites
\cite{Penc2011}.  This form shows why the $J_\mboti{bl}=J_\mboti{bq}$ point is $SU(3)$-symmetric:
only the first term above survives at this point.  This leads to the simplifying properties of this
special point which we have exploited in our discussion of the rung initial state.

\subsection{Mapping of the antisymmetric $1s$ subspace}\label{sup:AntiMap}
We describe now the mapping of the subspace of antisymmetric ($1s$) states of the two-leg Heisenberg
ladder with two excitations.  This mapping is exact and not perturbative.  The corresponding
subspace of the ladder is divided into a set $\mathcal{A}_1$ of states where the two excitations are
on different legs, and a set $\mathcal{A}_2$ where both excitations are on the same leg.  For each set,
we take the leg-antisymmetric combinations, since we are considering the antisymmetric subspace:
\begin{align}
    \ket{n_1, n_2}_{A_1} =& \frac{1}{\sqrt{2}} \left( \hat{S}_{n_1, 1}^- \hat{S}_{n_2, 2}^- - \hat{S}_{n_2, 1}^- \hat{S}_{n_1, 2}^- \right) \ket{0}_\mboti{ladder} \label{eq:AntiMap1} \\
    \ket{n_1, n_2}_{A_2} =& \frac{1}{\sqrt{2}} \left( \hat{S}_{n_1, 1}^- \hat{S}_{n_2, 1}^- - \hat{S}_{n_1, 2}^- \hat{S}_{n_2, 2}^- \right) \ket{0}_\mboti{ladder} \label{eq:AntiMap2}
\end{align}
when $\ket{0}_\mboti{ladder}$ denotes the ferromagnetic ground state of the ladder with all spins
up.  There are no matrix elements of the Hamiltonian mixing the two sets. One can check this by
noting that the only terms which could possibly connect the two sets are those exchanging spins on a
rung: $H^{xy}_\mboti{rung} = -J_y/2 \sum_r \{ \hat{S}_{r,1}^+ \hat{S}_{r,2}^- + \mbox{h.\,c.} \}$.
One finds by explicit evaluation that $H^{xy}_\mboti{rung} \ket{n_1, n_2}_{A_2}$ vanishes.  Hence
the subspaces $\mathcal{A}_1$ and $\mathcal{A}_2$ are disjoint.

Each of the two disjoint sectors is equivalent to a  spin-$\frac 12$ chain.  The translation is
defined by picking $n_1$ and $n_2$ as the positions of the overturned spins in the corresponding
chain: 
\begin{align*}
    \ket{n_1, n_2}_{A_1}\;  \leftrightarrow& \quad \hat{S}_{n_1}^- \hat{S}_{n_2}^- \ket{0}_\mboti{chain} \\
    \ket{n_1, n_2}_{A_2}\;  \leftrightarrow& \quad  \hat{S}_{n_1}^- \hat{S}_{n_2}^- \ket{0}_\mboti{chain}
\end{align*}
with $\ket{0}_\mboti{chain}$ meaning the ferromagnetic ground state of the spin-$\frac 12$ chain
(all spins in $+z$ direction).  The resulting chain Hamiltonians for both sectors are of the $XXZ$
type.  The parameters are different for the two sector  $\mathcal{A}_1$ and $\mathcal{A}_2$.  The
anisotropy is $\Delta_A=0$ for the  $\mathcal{A}_1$ sector and  $\Delta_A=1$ for the
$\mathcal{A}_2$ sector.  The other parameters and the form of the Hamiltonian is reported in the
main text, Sec.~\ref{sec:spectrum}.

As a side note: the mapping here can be generalized to multi-leg ladders with an even number of legs
and periodic boundary conditions in the $y$ direction. The cancellation of contributions from rung
exchange $H^{xy}_{\mboti{rung}}$ when acting on an $\mathcal{A}_2$ state holds pairwise for
neighboring legs of the more-leg ladder.  So the set of states
\begin{align*}
    \ket{n_1, n_2}_{\mathcal{A}_2} = \frac{1}{\sqrt{w}} \sum \limits_{y=1}^{w} S^-_{n_1, y} S^-_{n_2, y} \, \ket{0}_{\mboti{ladder}}
\end{align*}
is a disjoint subspace of the $w$-leg ladder and maps to the spin-$\frac 12$ chain with $\Delta_A =
1$.  Such a multi-leg ladder will also display a propagating mode that can be interpreted as a bound
magnon mode for this effective spin-$\frac 12$ chain, just as our two-leg ladder does. 

%% It is straightforward to generalize the statement above for multi-leg ladders with an even number of legs and periodic boundary conditions in the $y$ direction:
%% Since the Hamiltonian is a sum of two site operators, the cancellation of contributions from rung exchange $H^{xy}_{\mboti{rung}}$ when acting on an $\mathcal{A}_2$ state holds pairwise for neighboring legs of the more-leg ladder.
%% So the set of states
%% \begin{align*}
%%     \ket{n_1, n_2}_{\mathcal{A}_2} = \frac{1}{\sqrt{w}} \sum \limits_{y=1}^{w} S^-_{n_1, y} S^-_{n_2, y} \, \ket{0}_{\mboti{ladder}}
%% \end{align*}
%% builds a separated subspace of the $w$-leg ladder and maps to the spin-$\frac 12$ chain with $\Delta_A = 1$.
%% Since the eigenvalue of $H^z_{\mboti{rung}}$ when acting on an $\mathcal{A}_2$ state depends on the number of legs of the ladder, magnetic field $h_A$ and energy shift $\epsilon_A$ differ from the values given for the two-leg ladder above.
%% Preparing a multi leg system in the state $\ket{n, n+1}_{\mathcal{A}_2}$ and observing the time evolution of $\sum_{y=1}^{w} \langle S^z_{x, y}(t) \rangle/w$ shows the same propagation of a magnon bound state (2-string) as observed for the simple spin-$\frac 12$ chain when comparing to $\langle S^z_{x}(t) \rangle$.

\subsection{Mapping of the $2s$ space to the anisotropic spin-$\frac 12$ chain} \label{sup:MapD12}
According to Mila \cite{Mila1998} another mapping of the two-leg ladder this time acting on the singlet subspace can be performed.
A different first order approximation of the kind (\ref{eq:PerturbH}), App.~\ref{sup:BBMmap}, is obtained when $P$ projects on any combination of $\ket{s}$ and $\ket{t_{+}}$ states.
In order to receive a degenerate unperturbed rung subspace, we apply a magnetic field $h_c$
\begin{align*}
    H &= H_{\mbox{\tiny{rung}}} + h_c \sum_i (S_{i,1}^z + S_{i,2}^z) 
    + H_\mboti{leg} - h_c \sum_i (S_{i,1}^z + S_{i,2}^z) \\
      &= H_0 + H_1 \,,
\end{align*}
so for $h^c = J$, all ladder states built as arbitrary product states from $\{ \ket{s}, \ket{t_{+}} \}$ are degenerate with respect to $H_0$.
This time, we introduce the operators $S^\pm$ and $S^z$ acting on a chain (instead of a ladder) with
\begin{align*}
    \begin{array}{ll}
    S^+ \ket{s}  = \ket{t_{+}}, & S^+ \ket{t_{+}} = 0, \\
    S^- \ket{s}  = 0, & S^- \ket{t_{+}} = \ket{s}, \\
    S^z \ket{s}  = -\frac 12 \ket{s}, & S^z \ket{t_{+}} = \frac 12 \ket{t_{+}}\,.
    \end{array}
\end{align*}
By investigating the matrix elements of $H_1$ in this subspace, we get, up to a constant, an effective Hamiltonian describing an XXZ spin-$\frac 12$ chain with anisotropy $\Delta = 1/2$, magnetic field $h = -\frac 54 J_x$ and $J'' = J_x$
\begin{align*}
    H''_\mboti{eff} &= -\frac{J''}{2} \sum \limits_{\langle i,j\rangle} \left\{ S_i^+ S_{i+1}^- \mbox{h.\,c.} \right\}
    - J'' \Delta \sum \limits_{\langle i,j\rangle} S_{i}^z S_{i+1}^z \\
    &\qquad - h \sum \limits_{\langle i,j\rangle} S_i^z .
\end{align*}

\subsection{The Hilbert space at zero leg coupling $J_y = 0$} \label{sup:LowChiMap}
In order to present a complete picture of the Hilbert space, we discuss its structure at zero coupling of the legs of the ladder $\chi = 0$.
Similar to the antisymmetrized case discussed in App.~\ref{sup:AntiMap}, we divide the symmetrized subspace into states with one excitation per leg, building a basis set $\mathcal{B}_1$, and the remaining basis states consisting of two or zero excitations per leg $\mathcal{B}_2$.
An element of $\mathcal{B}_1$ is defined by the positions of the excitations $n_1$ and $n_2$
\begin{align*}
    \ket{n_1, n_2} &= \frac{1}{\sqrt{2}} \left( S^-_{n_1,1} S^-_{n_2,2} + S^-_{n_2,1} S^-_{n_1,2} \right) \ket{0},\quad n_1 < n_2 \\
    \ket{n_1, n_1} &= S^-_{n_1,1} S^-_{n_1,2} \ket{0}
\end{align*}
where $\ket{0}$ denotes the fully polarized state.
The ladder Hamiltonian $H$ is given by a sum of two site operators
\begin{align}
    \begin{split}
    H &= -\frac{J_x}{2} \sum \limits_{y=1}^2 \sum \limits_{\langle i,j \rangle_x} \left( S^+_{i,y} S^-_{j,y} + \mbox{h.\,c.} \right) \\
         &\quad -J_x \sum \limits_{y=1}^2 \sum \limits_{\langle i,j \rangle_x} S^z_{i,y} S^z_{j,y} \\
      &= \sum \limits_{\langle i,j \rangle_x} h^{+-}_{i,j} + \sum \limits_{\langle i, j \rangle_x} h^{zz}_{i,j} \,,
  \end{split}
  \label{eq:Huncoupled}
\end{align}
we focus on two-rung clusters and label each state by the magnetization of each rung $\ket{m_1, m_2}$, where $m_i=0,1,2$ defines the magnetization per rung as $m_i-1$.
Note that $\ket{1, 1}$ denotes $\ket{n_1=1, n_2=2}$ and is not a product state $\ket{1, 1} \neq \ket{1}_1 \otimes \ket{1}_2$.
In this basis, off diagonal matrix elements of $H$ are generated by $h^{+-}_{i,j}$.
As long as $j=i+1$, the matrix elements are independent of $i$ and we drop the indices, writing
\begin{align*}
    \bra{1, 0} h^{+-} \ket{0, 1} &= -\frac{J_x}{2} \\
    \bra{2, 0} h^{+-} \ket{1, 1} &= -\frac{J_x}{\sqrt{2}} \,.
\end{align*}
Instead of evaluating diagonal matrix elements of $H$ for the two site clusters individually, we note that
\begin{align*}
    \bra{n_1, n_2} J_x \sum \limits_{y=1}^2 \sum \limits_{\langle i,j \rangle_x} S^z_{i,y} S^z_{j,y} \ket{n_1, n_2} = \frac{J_x}{2} (L-4)
\end{align*}
for any choice of $n_1$ and $n_2$.
To compare these matrix elements to the non-interacting Bose Hubbard chain with two particle filling and a translation of Hilbert spaces where $n_1$ and $n_2$ describe the positions of particles in an otherwise empty chain $\ket{0}$
\begin{align*}
    \ket{n_1, n_2} &= a^\dagger_{n_1} a^\dagger_{n_2} \ket{0},\quad n_1 < n_2 \\
    \ket{n_1, n_1} &= \frac{1}{\sqrt{2}}\, a^\dagger_{n_1} a^\dagger_{n_1} \ket{0} \,,
\end{align*}
we regard the hopping term $h_{i,j}^{bh} = -\frac t2 a^\dagger_i a^{\phantom{\dagger}}_j +\, $h.\,c. again by considering two site clusters labeled by $m$, where this time $m$ defines the number of particles per lattice site, giving
\begin{align*}
    \bra{1, 0} h^{bh} \ket{0, 1} &= -\frac{t}{2} \\
    \bra{2, 0} h^{bh} \ket{1, 1} &= -\frac{t}{\sqrt{2}} \,.
\end{align*}
In the basis chosen, the hopping term has no diagonal matrix elements.
For the subspace $\mathcal{B}_1$, the Hamilton operators $H$ defined in Eq.~(\ref{eq:Huncoupled}) and $H_{bh}$ are equivalent, if $H_{bh}$ describes the non-interacting Bose Hubbard chain
\begin{align*}
    H_{bh} = \sum_{\langle i,j \rangle} h_{i,j}^{bh} - \mathbb{1}\, \frac{t}{2} (L-4)
\end{align*}
with $t=J_x$.

The remaining states $\mathcal{B}_2$ are further separated into states with two excitations on the first leg and a set of states with two excitations on the second leg.
These subsets are trivially equivalent to the spin-$\frac 12$ Heisenberg chain with two excitations and a constant energy shift caused by the empty leg.
There are no finite matrix elements of $H$ mixing states from separate subsets.
Hence $\mathcal{B}_2$ is described by the spin-$\frac 12$ Heisenberg chain.

\section{Spin current and Green's function} \label{sup:Current}
The form of a current operator which is supposed to display the flow of the magnetization in a spin
system depends on the Hamiltonian.  In this work, we consider different models leading to different
current observables.  Every system we study conserves the total magnetization which is expressed by
the continuity equation $\partial_t \langle S^z_i(t) \rangle = -[\mbox{div}\, j]_i$.  For a one
dimensional system, the divergence is given by the lattice version of Gauss law ($[\mbox{div}\, j]_i
= j_{i, i+1} - j_{i-1, i}$) and the magnetization current between sites $i$ and $i+1$ of the Heisenberg chain with coupling $J$
follows as
\begin{align*}
    j^{1'}_{i}(t)\, J = -J\, \mbox{Im} \, \langle S^+_i S^-_{i+1} \rangle \,.
\end{align*}
To be consistent with the main text, ``prime'' symbols are used to distinguish chain systems from ladder systems (no prime).
For the bilinear biquadratic model (BLBQ) we find
\begin{align*}
    j_{i} &= -J_\mboti{bl}\ \mbox{Im}\, \langle S^+_i S^-_{i+1}  \rangle\\
            &\quad-J_\mboti{bq}\ \mbox{Im} \, \langle S^+_i S^-_{i+1} {\bf S}_i \cdot {\bf S}_{i+1}
            + {\bf S}_i \cdot {\bf S}_{i+1} S^+_i S^-_{i+1}  \rangle\\
            &= -J_\mboti{bl}\ \mbox{Im}\, \langle S^+_i S^-_{i+1}  \rangle\\
            &\quad -\frac{J_\mboti{bq}}{2}\ \mbox{Im} \, \langle S^+_i S^+_i S^-_{i+1} S^-_{i+1}  \rangle\\
            &\quad -J_\mboti{bq}\ \mbox{Im} \,\langle S^z_i S^z_{i+1} S^+_{i} S^-_{i+1}  \rangle\\
            &\quad -J_\mboti{bq}\ \mbox{Im} \,\langle S^+_{i} S^-_{i+1} S^z_i S^z_{i+1}  \rangle
\end{align*}
what can be shown by evaluating the commutator $[S^z_i,H]$ and using the expression 
\begin{align*}
    \left[ S_i^z, {\bf S}_j \cdot {\bf S}_{j+1} \right] &= \frac{1}{2} \left[ \delta_{i,j} ( S_i^+ S_{i+1}^- - S_{i}^- S_{i+1}^+ ) \right. \\
    &\quad \left. - \delta_{i,j+1} ( S_{i-1}^+ S_{i}^- - S_{i-1}^- S_{i}^+ ) \right] \,,
\end{align*}
which leads to the divergence of the magnetization current as defined by the continuity equation
\begin{align*}
[\mbox{div}\, j]_i = 
     i \frac{J_\mboti{bl}}{2} &\left( [ S_i^+ S_{i+1}^- -S_{i-1}^+ S_i^-] - \mbox{h. c.} \right) \\
     + i \frac{J_\mboti{bq}}{2} &\left( {\bf S}_i \cdot {\bf S}_{i+1} [ S_i^+ S_{i+1}^- - \mbox{h.c.}] \right. \\
     &\quad \left. + [ S_i^+ S_{i+1}^- - \mbox{h.c.}] {\bf S}_i \cdot {\bf S}_{i+1}
      - ["i \rightarrow i-1"] \right) \,.
\end{align*}
The current of doubly occupied sites $S^z= -1$ however has to be defined in a different way as the
projector $P^\downarrow_i = S^z_i(S^z_i-1)/2$ is not conserved and does not fulfill a continuity
equation.  To analyze the movement of $\ket{t_-}$ triplets, we focus on the four point Green's
function (note the change of sign for tracking the movement of down spins instead of up spins)
\begin{align*}
    j^{2'}_{i}(t) = \mbox{Im}\ \langle S_i^+ S^+_i S^-_{i+1} S^-_{i+1} \rangle
\end{align*}
which also contributes to the magnetization current of the BLBQ .

We would like to emphasize, that the physical magnetization current is an observable which depends on the system under consideration, but the observables $j^{1'}_i(t)$ and $j^{2'}_i(t)$ describe the overlap of wave functions with adjacent occupation.
It is therefore always connected to the movement of single excitations or rung excitations, although the interpretation as a current is only valid for particular Hamiltonians.
We choose $j_i^{1'}(t)$ and $j^{2'}_i(t)$ to track direction dependent movement in all spin-1 systems we discuss.

\section{Broadening of the dispersion associated with the rung initial state according to Fermi's golden rule}\label{sup:Fermi}
In this Section we derive an estimate of decay rates of the rung initial state by using Fermi's golden rule.
We focus on the case of two excitations in the $SU$(3)-permutation model $J_{\mboti{bl}} = J_{\mboti{bq}}$.
States where both particles are on the same site correspond to a polarized $S^z=+1$ chain with a single overturned site $S^z=-1$.
In the permutation model, these states are separated from the rest of the Hilbert space (where both excitations are on different sites) and the eigenstates are simple magnons
\begin{align*}
    \ket{\psi_{k}} = \frac{1}{2 \sqrt{L}}\, \sum \limits_{n=1}^L e^{i k n}\,S^-_n S^-_n \ket{0} \,.
\end{align*}
On the other hand, the disjoint subspace defined by two particles on different sites is equivalent to the spin-$\frac 12$ Heisenberg chain and therefore analytically solvable as well.
The subspace is divided into $L-3$ bound states and $L(L-3)/2 + 3$ additional states~\cite{Karabach1997}.
In the continuum limit $L\rightarrow \infty$, these additional states become two superimposed magnons
\begin{align}
   \ket{\gamma_{k_1 k_2}} \sim \sum \limits_{\ell<\ell'}
   \{e^{i (k_1 \ell + k_2 \ell')} + e^{i (k_1 \ell' + k_2 \ell)}\} S^-_\ell S^-_{\ell'} \ket{0} \,.
   \label{eq:ScatState}
\end{align}
We are interested in the decay of rung states into these scattering states hence we ignore the bound states for now.
In order to estimate the decay width of $\ket{\psi_{k}}$-states, when $J_\mboti{bl} - J_\mboti{bq}$ is small but finite, we consider Fermi's golden rule meaning a second order perturbative treatment
\begin{align}
    \Gamma = \left. 2 \pi\, |V_{mi}|^2 \rho \right|_{E_i \sim E_m} \,,
    \label{eq:FGR}
\end{align}
where $i$ denotes some rung state and $m$ means a scattering state~\cite{Sakurai1993}.
Here, we get
\begin{align*}
    %V_{mi} = -4 (J_{\mboti{bl}} - J_{\mboti{bq}})\, \frac{J_{\mboti{bq}}}{J_{\mboti{bl}}}\, (1 + \cos{k})\,.
    V_{mi} = -4 (J_{\mboti{bl}} - J_{\mboti{bq}})\, ( \cos{k_1} + \cos{k_2} ) \delta_{k, k_1 + k_2}
\end{align*}
The density of states $\rho$ is obtained from the energy of the scattering states
\begin{align}
    E_\gamma(k_1,k_2) &= 2 J_{\mboti{bl}}( 2 - \cos{k_1} - \cos{k_2} ) , \quad k_2 = k - k_1 \notag \\
    \Rightarrow \frac{1}{\rho} &= \frac{\partial E}{\partial k_1} \\
    &= 2 J_{\mboti{bl}}\, \left( (1+\cos{k})\, \sin{k_1} - \sin{k}\, \cos{k_1} \right) \,.
    \label{eq:dos}
\end{align}
The matrix element has to be evaluated between states $\ket{\psi_{k}}$ and $\ket{\gamma_{k_1 k_2}}$ which share the same energy $E_{k}$.
The same holds for the density of states which is also evaluated at $E_{k}$, meaning that 
\begin{align}
    E_{\gamma}|_{k} = E_{\psi}|_{k} = 4 J_{\mboti{bl}} - 2 J_{\mboti{bq}} (1 + \cos{k}) \,,
    \label{eq:EqualEnergy}
\end{align}
see (\ref{eq:rung_energy}), Sec.~\ref{sec:BLBQ}, which allows us to express $\cos{k_1}$ as a function of $k$
\begin{align*}
    \cos{k_1} &= \frac 12 \left( \frac{J_{\mboti{bq}}}{J_{\mboti{bl}}}\, (1 + \cos{k}) \right. \\
    &\quad \left. + \sqrt{(1-\cos{k})\, (2 - (\frac{J_{\mboti{bq}}^2}{J_{\mboti{bl}}^2})(1+\cos{k}) )} \right) \,.
\end{align*}
Combining all these conditions into Fermi's golden rule, we receive a reasonable decay width as shown in Fig.~\ref{fig:SemiAnalytics}.
A comparison of analytical and numerical results confirms that eigenstates of the rung sector of the $SU(3)$-permutation model around $k \approx \pi$ are stable despite the perturbation, though the lifetime of these states decreases with $k$.

\section{Connection of dispersion and wave front dynamics} \label{sup:EnginDyn}
A correspondance we used throughout the discussion of the ladder dynamics is given by the connection
of dispersion $\epsilon(k)$ and propagating wave fronts.  To be precise, our intial states are all
spatially localized, static states, hence they are built from an equal sum over all Fourier
transformed states, i.\,e. the whole Brillouin zone (BZ),
\begin{align*}
    \ket{\psi} = \frac{1}{\sqrt{L}} \sum_{k \in BZ} c_k \ket{\psi_k} ,\quad c_k = 1\ \forall k \,.
\end{align*}
If all $\ket{\psi_k}$ happen to be eigenstates (or a sum of eigenstates close in energy) of the
system, the spectral decomposition shows a single (broadened) peak for each value of $k$.  In this
case, a dispersion, relating momenta $k$ to energies $\epsilon(k)$ can be defined.  A statement
about the corresponding real space dynamics is then given as follows: If $\epsilon(k)$ is two times
differentiable in the BZ and has an inflection point $k^*$ in this interval, then the real space
dynamics is expected to show the propagation of wave fronts with a velocity $v$, given by the group
velocity $v_g = \partial \epsilon(k) / \partial k$ at the inflection point $v = v_g(k^*)$.

As has been discussed by~\cite{Ganahl2012a}, the reason for this is that the density of
states as a function of $v_g$ diverges at the inflection point $v_g(k^*)$.
\begin{align*}
    \rho(v_g) &= \int_k \delta(v_g - v(k)) \\
    &= \frac{1}{2 \pi}\, \int \frac{1}{v_g'}\ \delta(v_g - v) \mbox dv \\
    &= \frac{1}{2 \pi\, v_g'(k)} \,.
\end{align*}
Since $v_g'(k^*) = 0$, the density of states diverges at $k = k^*$.  Although we considered the
dispersion to be defined for the whole BZ, our numerical analysis of the ladder dynamics suggests
that this correspondance is valid for weakened cases, e.\,g. if the domain of $\epsilon(k)$ is
diminished $k \in [k_0, 2 \pi - k_0], k_0 \in (0, \pi)$.

There is a simple interpretation of the argument.  Localized states are composed of sums of momentum
states, where every momentum has its contribution, e.\,g. $\ket{\psi} = \sum_k \ket{\psi_k}$.
Regarding the group velocity, we assign a velocity to every momentum state by considering the
dispersion $\epsilon(k)$.  Around the inflection point of the dispersion, the change of velocity
from a small change of the momentum has a minimum.  Therefore, most contributions to the group
velocity are collected from velocities around this inflection point of $\epsilon(k)$ (which
translates to an extremum of the group velocity $\partial \epsilon(k)/ \partial k$ if $\epsilon(k)$
is defined for $k \in [0, 2 \pi)$\,).

\section{Absence of bound-state wave front for $\Delta=0.5$} \label{sup:no_bound_state_at_small_Delta}
We noted that no visible wave front is associated with the $2s$ sector, which can be mapped to a
spin-$\frac 12$ chain XXZ chain with $S^z$-anisotropy $\Delta = \frac 12$.

Here, we discuss the vanishing of wave fronts when decreasing $\Delta=1$ to $\Delta=0.5$ and
describe what happens for the corresponding spectral decompositions.  The propagation of $m$-strings
in the spin-$\frac 12$ XXZ chain at various $\Delta$ has been studied, both numerically and via
Bethe ansatz, in previous work~\cite{Ganahl2012a}.  We review the loss of wave
fronts observed for the two-string propagation for decreasing $\Delta$.

Since the problem is symmetric in the Brillouin zone (BZ) and mirrored at $k=\pi$, we focus for our
description on $k \in [0, \pi]$.  The scattering continuum of the two-particle sector of the
spin-$\frac 12$ XXZ chain is given by
\begin{align*}
    \epsilon_{\mbt{cont}}(k_1, k_2, \Delta) = J_x (2 \Delta - \cos{k_1} - \cos{k_2})\,,
\end{align*}
with $k_1, k_2 \in [0, 2\pi)$ \cite{Karabach1997}.
The lower edge of the scattering continuum is defined for $k_1 = k_2 = \frac{k}{2}$, hence $\epsilon_{\mbt{edge}}(k, \Delta) = \epsilon_{\mbt{cont}}(k/2, k/2, \Delta)$.
As a second ingredient, the dispersion of the two-string is given by 
\begin{align*}
    \epsilon_2(k, \Delta) = \frac{J}{2 \Delta} ( 2\Delta^2 - 1 - \cos{k} )\,.
\end{align*}
The inflection point of $\epsilon_2(k,\Delta)$ is found to be $k^* = \frac{\pi}{2}$.  According to
the mechanism given by Eqn.~\ref{eq:DosArg} and described in Sec.~\ref{sup:EnginDyn}, two-strings
propagate as wave fronts with a velocity $v_2$, equal to the group velocity $v_g = \partial
\epsilon_2(k, \Delta)/ \partial k$ at $k^*$, giving $v_2(\Delta) = J/(2 \Delta)$.  As mentioned in
the main text, for $\Delta = 1$, there is no crossing point of $\epsilon_{\mbt{edge}}$ and
$\epsilon_2$.  If $\Delta < 1$, a crossing point exists and is given by $k'(\Delta) = 2
\arccos{\Delta}$.  The two-string disperion $\epsilon_2$ is separated from the lower edge of the
scattering continuum $\epsilon_{\mbt{edge}}$, only for a part of the BZ $k \in [k', \pi]$ (remember
that we restricted our description to $k \in [0, \pi]$ because of the symmetry of the problem).  At
$\Delta = \Delta_c = 1/\sqrt{2}$, the crossing point reaches the inflection point of $\epsilon_2$,
meaning $k'(\Delta_c) = k^*$ so an inflection point is not defined for $\Delta \leq \Delta_c$.  The
absence of an inflection point results in the absence of propagating wave fronts for the real time
evolution (App.~\ref{sup:EnginDyn}).  Furthermore, we note that this connection of dispersion and
real space dynamics may be formulated like ``wave fronts propagate with a velocity, given by the
maximum group velocity, i.\,e. the maximum derivative of the dispersion'', instead of our
forumlation via inflection points.  However, this alternative formulation is only true if the
maximum group velocity not realized at the edges of the domain of the dispersion.  In other words,
the inflection point of $\epsilon_2$ becomes ill-defined upon crossing $\Delta_c$ from $\Delta >
\Delta_c$ to $\Delta < \Delta_c$; a maximum group velocity on the other hand is defined for any
$\Delta \in (0, 1]$.

\begin{figure} 
    \centering 
    \includegraphics[width=0.5\textwidth]{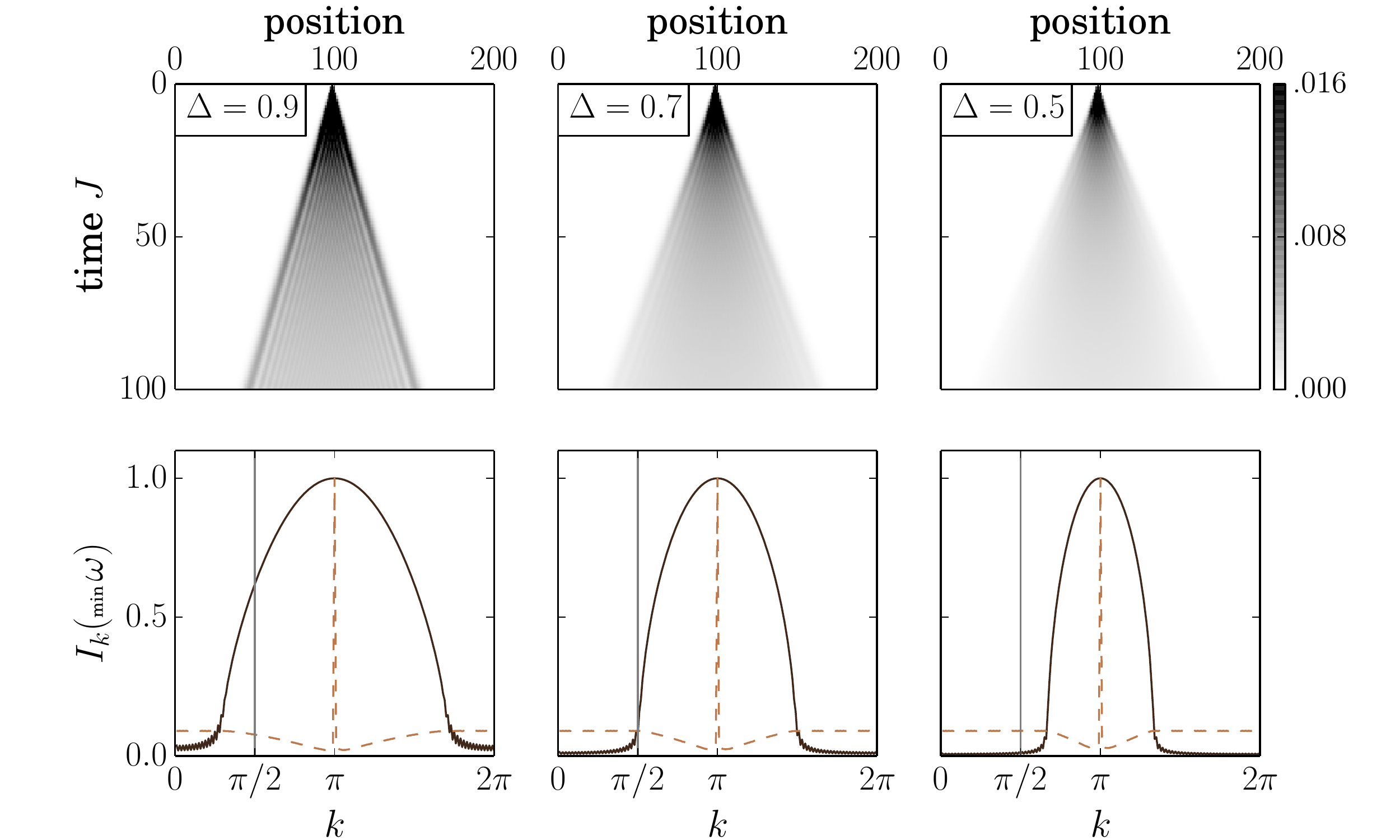}
    \caption{ Dynamics and spectral properties of two initially neighboring excitations in the spin-$\frac 12$ XXZ chain for different values of the $S^z$-anisotropy $\Delta=0.9, 0.7, 0.5$ from left through right column (the case $\Delta=0.5$ (right column) is equivalent to the $2s$ sector).
        First line: Time evolution of the projection on neighboring excitations in the spin-$\frac 12$ XXZ chain $P^{\downarrow \downarrow}_i(t)$.
        Second line: Solid lines display the overlap with states building the lower edge of the energy spectrum, dashed lines denote the average weight per pole;
        numerical data for a system with $L=200$ lattice sites. Note that the weight of the lowest poles at the edges of the Brillouin zone will scale to zero for larger system sizes.
        Comparing the weight of the lowest poles of the spectral decompositions to the average pole weight allows to determine a momentum $k_0$ for which the lower branch of the spectrum dissolves into the scattering continuum.
        We note that propagating wave fronts seem to be lost when $k_0 > \pi/2$ and the remaining dispersion $\epsilon_k = J(1-\cos{k})/2$ for $k \in [k_0, 2 \pi-k_0]$ has no linear part anymore, i.\,e. a maximum group velocity is ill-defined.
        This is clearly the case for the $2s$ sector hence no dominant wave fronts are observed besides the ones corresponding to the trivial mode of single particle propagation (not visible for $P_i^{\downarrow \downarrow}(t)$).
    }
    \label{fig:HiddenMode}
\end{figure}

In order to estimate $k_0$ from our numerical data, we compute the weight of the lowest pole for
each momentum and compare it to the average pole weight, Fig.~\ref{fig:HiddenMode}.  The dominance
of the lowest poles is lost when the pole weight is of the order of the average pole weight,
defining the points where the lower branch of the energy spectrum ``enters'' the scattering
continuum.  We note that $k_0 \sim \pi/2$ for $\Delta \sim 0.7$, defining the regime where dominant
wave fronts become absent and we do not observe them anymore for the $2s$ sector $(\Delta=0.5)$.

%%%%%% FIGURE %%%%%% FIGURE %%%%%% FIGURE %%%%%% FIGURE %%%%%% FIGURE %%%%%%%%%%
%%%%%% FIGURE %%%%%% FIGURE %%%%%% FIGURE %%%%%% FIGURE %%%%%% FIGURE %%%%%%%%%%

\section{Technical remarks on the characterization of non-ballistic spreading}\label{sup:TechRem}
%%%%%% FIGURE %%%%%% FIGURE %%%%%% FIGURE %%%%%% FIGURE %%%%%% FIGURE %%%%%%%%%%
\begin{figure} 
    \centering 
    \includegraphics[width=0.5\textwidth]{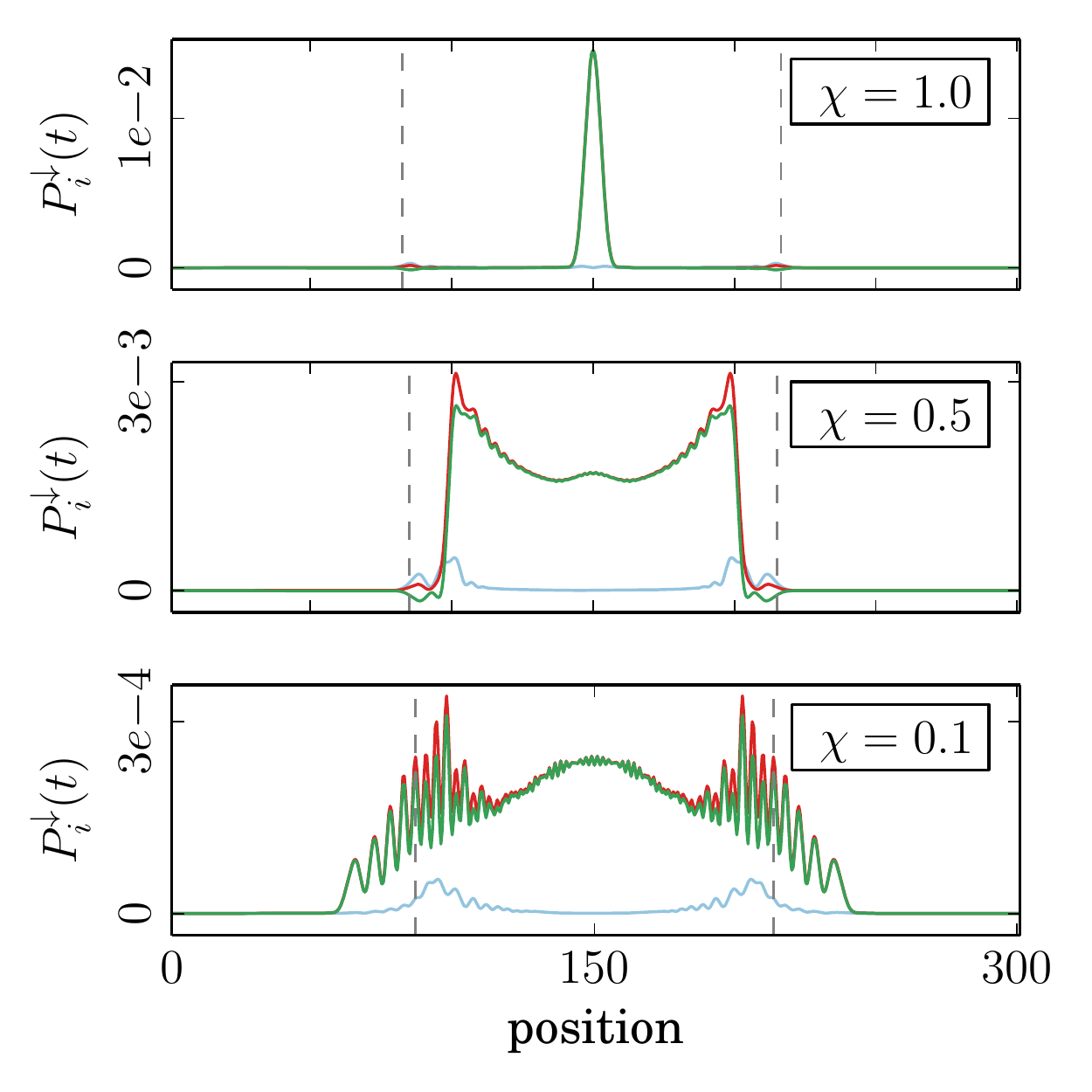}
    \caption{
        Snapshots at time $t=100/J_x$, rung initial state, three $\chi$ values.  Each plot shows three different observables: projection on two excitations sharing a rung $P^\downarrow_i(t)$ (red lines), projection on two excitations neighboring on a leg $P^{\downarrow \downarrow}_i(t)$ (blue lines) and $P^\downarrow_i(t) - P^{\downarrow \downarrow}_{i-\frac 12}(t)$ (green lines). A 5-point moving average filtering has been performed on all observables to damp out peaks.  The fast wave front corresponding to the triplet mode is marked by vertical dashed lines; at large $\chi$ this is clearly distinguished from the rung-specific mode.  For small $\chi$ the two modes of propagation can not be separated by our procedure because they have similar expansion velocities and interfere heavily.} 
    \label{fig:RungSigma_profiles}
\end{figure}
%%%%%% FIGURE %%%%%% FIGURE %%%%%% FIGURE %%%%%% FIGURE %%%%%% FIGURE %%%%%%%%%%
%%%%%% FIGURE %%%%%% FIGURE %%%%%% FIGURE %%%%%% FIGURE %%%%%% FIGURE %%%%%%%%%%
\begin{figure} 
    \centering 
    \includegraphics[width=0.5\textwidth]{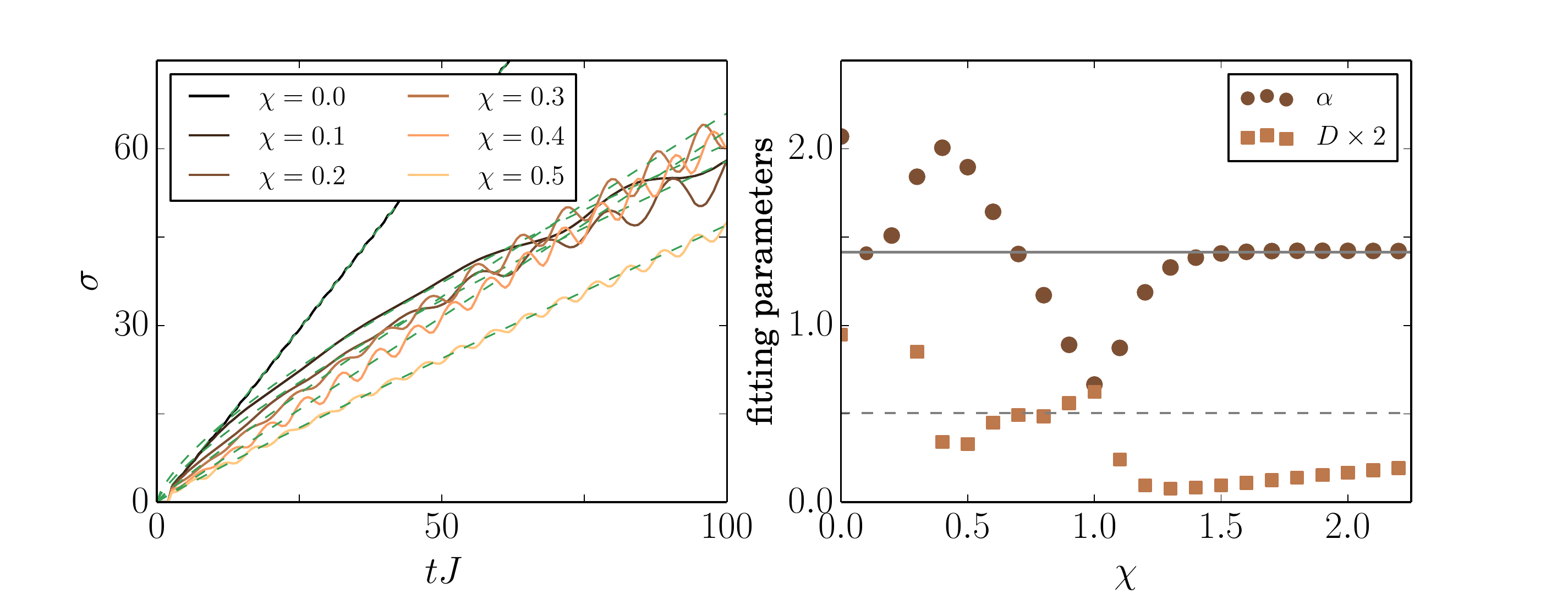}
    \caption{Figure corresponding to Fig.~\ref{fig:Sigma}, showing the behavior of $\sigma(t)$ for the dynamics of the rung initial state with a focus on the low-coupling regime and a system with $2L = 602$ lattice sites.
    Green dashed lines denote applied fits $\sigma^2(t) = D\,t^\alpha$.} 
    \label{fig:RungSigma_LowChi}
\end{figure}
%%%%%% FIGURE %%%%%% FIGURE %%%%%% FIGURE %%%%%% FIGURE %%%%%% FIGURE %%%%%%%%%%
In Sec.~\ref{sec:RungAnalysis}, a discussion of the rung dynamics and its specific mode showing non-ballistic spreading has been presented.
Here, we provide additional information on the numerical evaluation of this dynamic feature.
The most distinctive signal for this mode of propagation is found for the projection on rung states $P^\downarrow_i(t)$ which we used to determine the spatial standard deviation $\sigma$ defined in \eqref{eq:sigma_def}.
As a technical remark, we note that the rung initial state weakly overlaps with the bound state branch associated with the triplet mode discussed in~\ref{sec:TripMode}) (see for instance Fig.~\ref{fig:MainRung}, third line).
While for the leg intial state the triplet mode of propagation was a dominant feature, it is hardly visible for the dynamics of the rung initial state, especially when considering the obersvable $P^\downarrow_i(t)$.
In Fig.~\ref{fig:RungSigma_profiles}, we display the observable for exemplaric coupling strengths representing the low, the intermediate, and the high coupling regime at a specific time $t=100/J_x$.
Our analysis focusses on the slow, central peak observed for $\chi \gtrsim 0.5$.
In order to exclude deteriorating effects from the weak signal associated with the triplet mode, we modify the distribution function $P^{\downarrow}_i(t)$ by subtracting the projection on excitations neighboring on the same leg $P^{\downarrow \downarrow}_i(t) = \sum_y \langle (S^z_{i,y} -\frac 12)(S^z_{i+1,y} -\frac 12)\rangle$.
We put negative values of $P^{\downarrow}_i - P^{\downarrow \downarrow}_i$ to zero and end up with a cleaner distribution function for the description of the rung-specific mode of propagation.
Note that $P^{\downarrow \downarrow}_i(t)$ is actually defined on bonds between $i$ and $i+1$, hence we shift the observable by half a lattice site before subtraction.

In Fig.~\ref{fig:RungSigma_LowChi}, the behavior of the width characterizing parameter $\alpha$, according to our method, is displayed for all regimes of coupling strengths.
For $\chi=0$, $P^{\downarrow\downarrow}_i(t)$ is exactly zero and $P^\downarrow_i(t)$ shows ballistic behavior $\alpha \sim 2$.
Especially for the regime where the rung-specific mode of propagation assumes spreading via wave fronts $\chi < 0.5$, the triplet mode and the rung-specific (resonance) mode interfere and the exponent $\alpha$ determined with our method is expected to characterize neither of both modes of propagation.
A deeper analysis of the spatial standard deviation in the low coupling regime is left for future work.

\section{Observables in the uncoupled limit}\label{sup:LowChi}
%%%%%% FIGURE %%%%%% FIGURE %%%%%% FIGURE %%%%%% FIGURE %%%%%% FIGURE %%%%%%%%%%
\begin{figure*} 
    \centering 
    \includegraphics[width=0.3\textwidth]{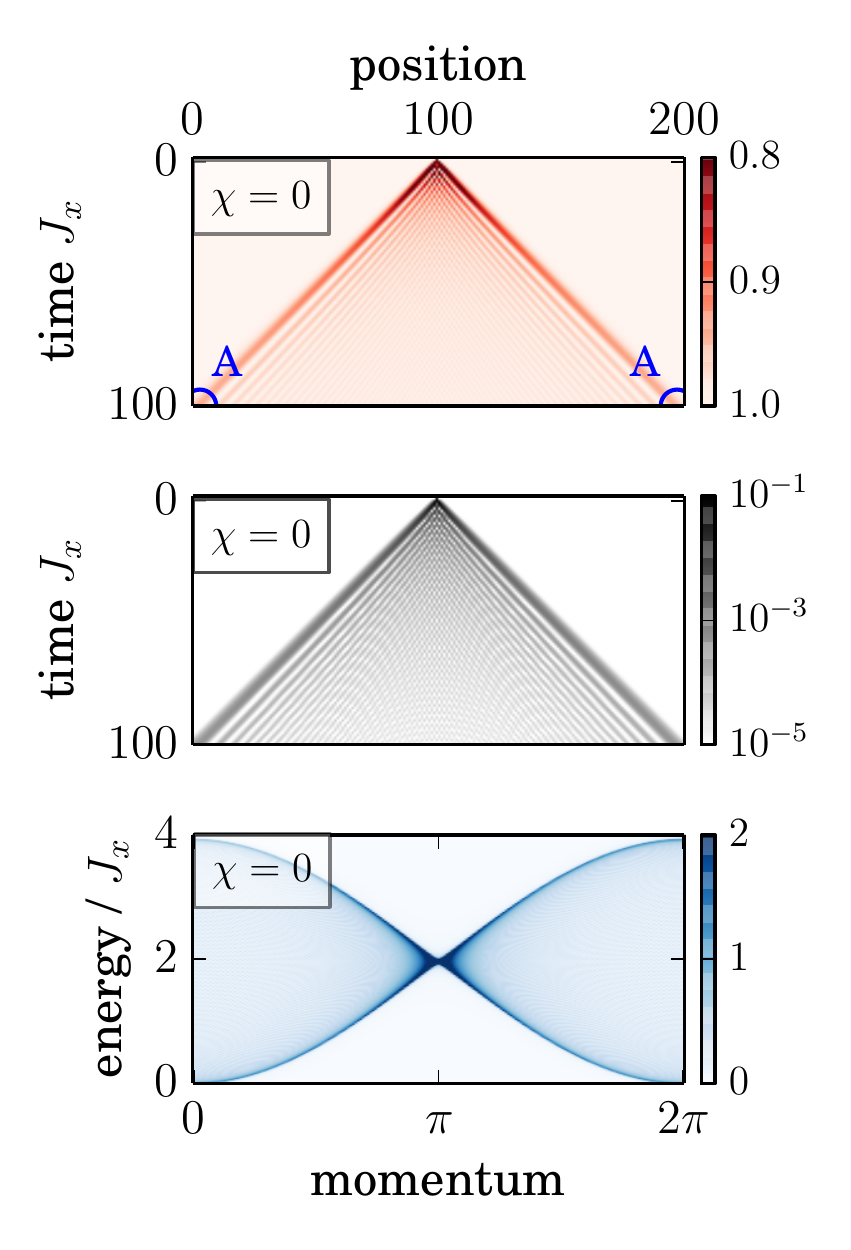}
    \includegraphics[width=0.3\textwidth]{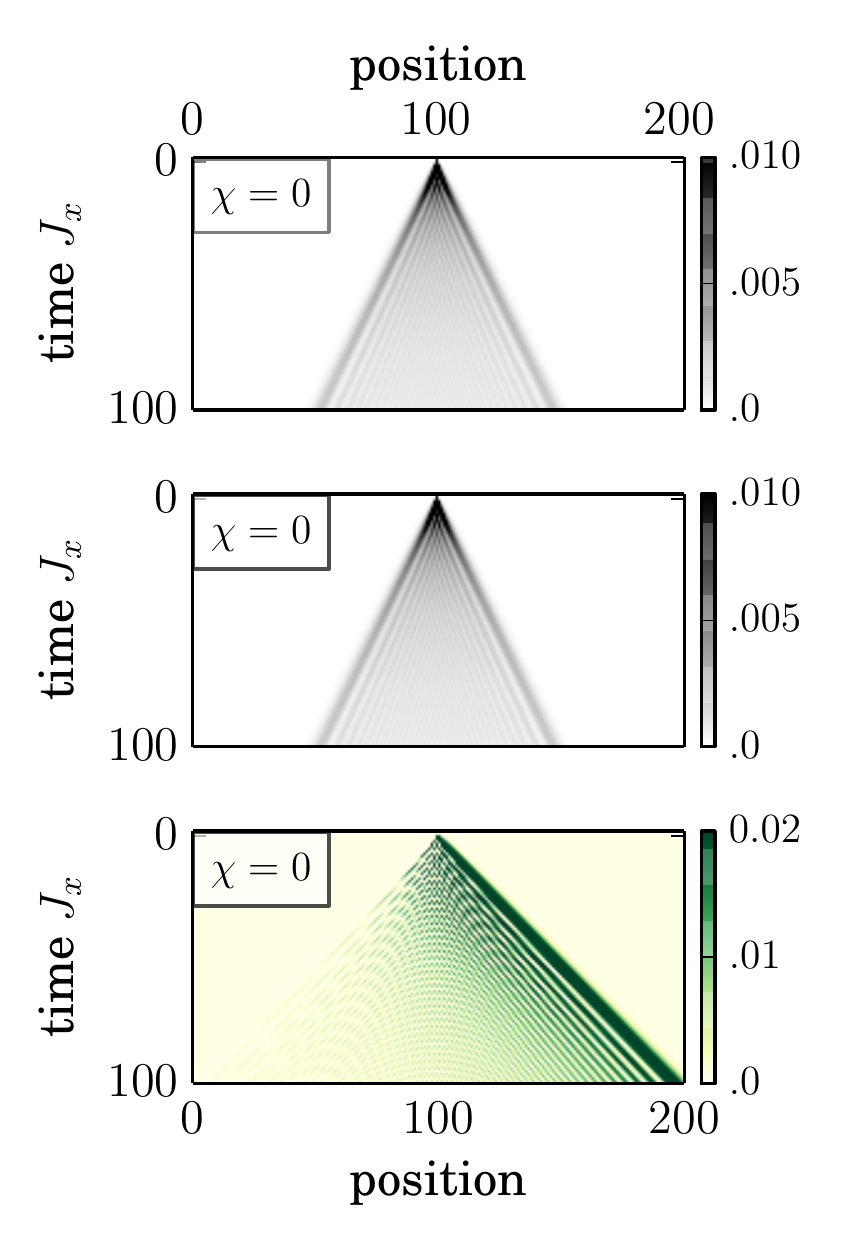}
    \includegraphics[width=0.3\textwidth]{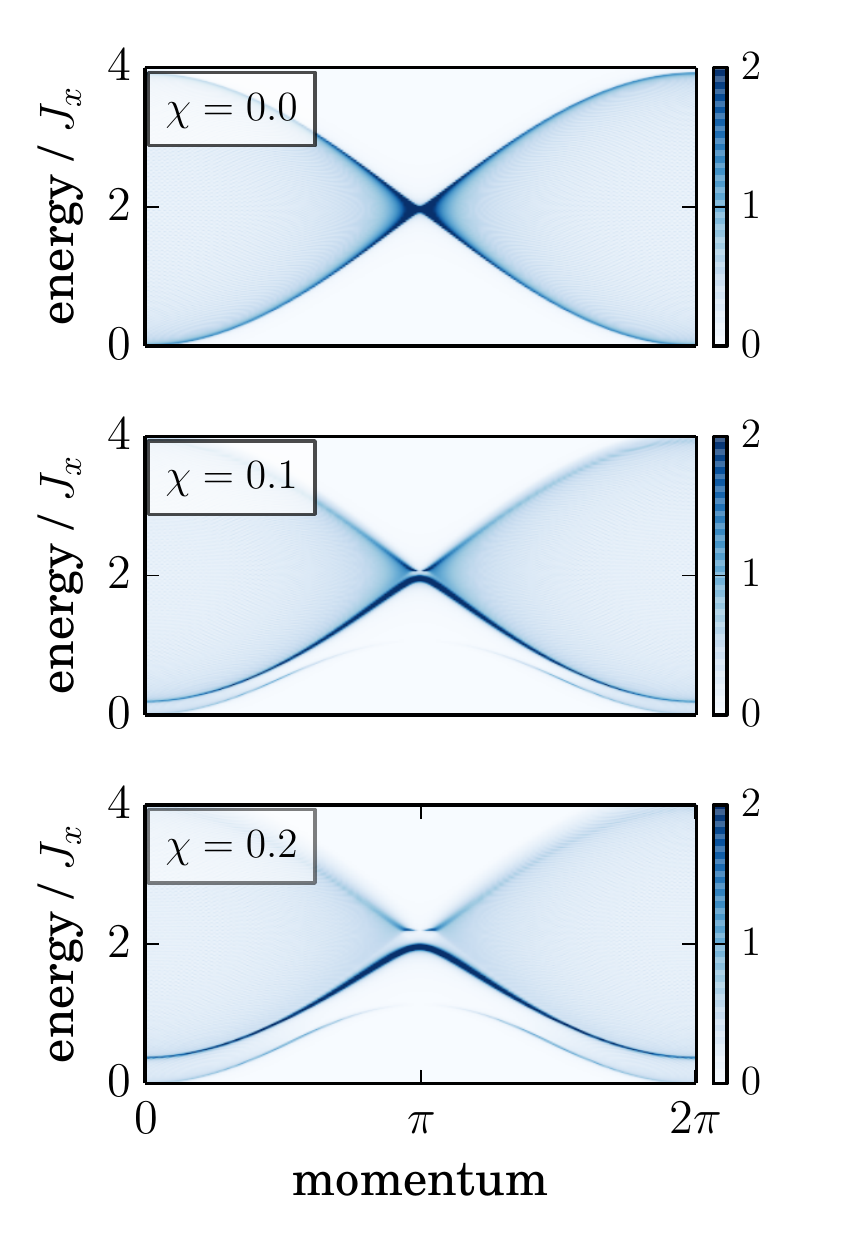}
    \caption{Selection of key figures in the limit $\chi=0$ and systems with $2L=200$ lattice sites.
        The left column corresponds to Fig.~\ref{fig:MainRung}, i.e. magnetization, rung projection and spectral decomposition (from first to third line) for the rung initial state.
        In the second column, we display projection on neighboring triplets (top) and singlets (center) of the leg initial state. These figures correspond to Fig.~\ref{fig:SymLegDyn}.
        The lowermost panel of the second column displays the current function $j^1_i(t)$ for the rung initial state (see Fig.~\ref{fig:Curr}).
        In the right column, the formation of the resonance as observed for the rung initial state is shown for growing values of the leg coupling $\chi$.
}
\label{fig:GenAddInfo}
\end{figure*}
%%%%%% FIGURE %%%%%% FIGURE %%%%%% FIGURE %%%%%% FIGURE %%%%%% FIGURE %%%%%%%%%%
To present a more complete picture we report on the behavior of our observables in the chain limit $\chi = 0$.
In this limit, the leg initial state exactly assumes all properties from the simple spin-$\frac 12$ chain with two initially neighboring excitations.
Again, we consider the division of local basis states into states with exactly one particle per leg $\mathcal{B}_1$, and remaining basis states $\mathcal{B}_2$, see App.~\ref{sup:LowChiMap}.
Projection operators projecting on these two sets are given by $P_1$ and $P_2$ respectively.
Since for $\chi=0$ the leg initial state is confined to $\mathcal{B}_2$ (meaning $P_1 \ket{\mbox{leg(t)}} = 0$), and the projection of neighboring triplets and singlets are identical ($P_2 \ket{t_0 t_0} = P_2 \ket{ss}$, see Fig.~\ref{fig:SymLegDyn}), both observables $\bra{t_0 t_0}\psi(t)\rangle$ and $\bra{ss}\psi(t)\rangle$ are equivalent, see Fig.~\ref{fig:GenAddInfo}.

The rung initial state in the low-$\chi$ limit is more peculiar.
A resonance associated with the rung-specific mode of propagation is only present for a non-zero rung coupling.
For $\chi=0$, dynamics of the rung initial state is described by two non-interacting chains carrying a single excitation.
The spectral decomposition of the rung initial state does not show a dispersion corresponding to a resonance branch.
This branch quickly develops for fininte coupling of the legs of the ladder, see Fig.~\ref{fig:GenAddInfo}~right column, and forms an associated mode of propagation, see Fig.~\ref{fig:GenAddInfo}~left column.
Note that this resonance forms at the edges of the scattering continuum of the spectrum, hence shows a larger slope and associated velocity than the triplet mode of propagation.
The triplet mode of propagation overlaps with the lower bound state branch which approaches the simple 2-string dispersion in the low coupling limit.
In Fig.~\ref{fig:GenAddInfo}, we furthermore display the current $j^1_i(t)$ (the Green's function $j^2_i(t)$ is exactly zero in the uncoupled limit), as well as magnetization and rung projection $P_i^\downarrow(t)$ for the uncoupled limit.

\section{Numerics - Obtaining spectral functions}\label{sup:SpecFcts}
To study the time evolution of a specific initial state, we project the propagator to a Krylov subspace generated from the initial state and the Hamiltonian by applying the Lanczos algorithm.
A sufficient subspace is generated iteratively and usually much smaller than the Hilbert space.
This method enables a diagonalization of systems with sizes way beyond the limit reachable with full (e.g. Cholesky) diagonalization~\cite{Park1986}.
To further increase the number of lattice sites of our systems, we exploit the fact that the Heisenberg model conserves the number of excitations (i.e. spin flips with respect to the ferromagnetically ordered background state).
Since all initial states we consider have a well-defined magnetization, the dimension of the relevant subspace grows quadratically in system size ${2 L}\choose{2}$.
Performance of the algorithm generally depends on the initial state and properties of the Hamiltonian.
On an ordinary workstation, we were able to simulate system sizes of up to a thousand rungs (full diagonalization algorithms are roughly limited to a few hundreds of rungs).

An advantage of the Lanczos method we use to perform the time evolution is the possibility to
extract dynamical response directly from the tridiagonal matrix $T_m$, with $\alpha_i$ as diagonal
and $\beta_i$ as off diagonal entries, obtained during the computation~\cite{Gagliano1987}.  In particular, we are
interested in the spectral function
\begin{align*}
    I(\omega) = \sum \limits_{j} |\bra{v_j} \psi \rangle |^2\, \delta(\omega - (E_n-E_0)) \,,
\end{align*}
which displays the composition of some state $\ket{\psi}$ in terms of eigenstates $\ket{v_j}$ with energy $\omega$.
Using the Cauchy principal value, the expression is equivalent to
\begin{align*}
    I(\omega) = - \frac{1}{\pi} \mbox{Im} [\bra{\psi} \frac{1}{\omega + E_0 + i \epsilon - \hat{H}} \ket{\psi}] \,.
\end{align*}
Here, $E_0$ is the ground state energy and $\epsilon$ is a small real number to shift the poles into the complex plain.
For our numerical computation of $I(\omega)$, $\epsilon$ is a cosmetical parameter to artificially broaden peaks and make them plotable, although the choice of $\epsilon$ also scales the values of $I(\omega)$.
All spectral decompositions we show are obtained with a fixed $\epsilon=0.01 J_x$ for the ladder and $\epsilon=0.04$ for chain systems. 
A smaller value is chosen for the spin ladder since the Hilbert space has a higher density when compared to chain systems and we demand a more refined picture for these cases.
The number of Lanczos iterations is chosen to be of the order of $\sim 1000$, which is a sufficiently large value such that results do not change for larger choices of this number.

Our goal is to compute the quantity
\begin{align*}
    x_0 = \bra{\psi} \frac{1}{z - \hat{H}} \ket{\psi} \,,
\end{align*}
with $z = \omega + E_0 +i \epsilon$.
Therefore, we follow an approach presented in \cite{Fulde1991} and begin with
\begin{align}
    (z-\hat{H}) (z - \hat{H})^{-1} &= \mathbb{1} \nonumber \\
    \sum \limits_j \bra{v_m} (z - \hat{H}) \ket{v_j} \bra{v_j} (z-\hat{H})^{-1} \ket{v_n} &= \delta_{m,n}\,. \label{eq:LanApp}
\end{align}
We obtain a system of linear equations $\sum_j (z-\hat{H})_{m,j} x_j = \delta_{m,1}$ when we define $x_j = (z-\hat{H})^{-1}_{j,1}$.
The solution is formally given by Cramer's rule $x_j = $~det$(\hat{M}_j)/$~det$(z-\hat{H})$ while $\hat{M}_j$ is identical to $z-\hat{H}$ when the $j$th column is modified according to $\hat{M}_{j,n} = \delta_{n,1}$.
Of course we already applied the Lanczos approximation in step (\ref{eq:LanApp}) by assuming that the Krylov space generated during the Lanczos iteration is complete.
The matrix $(z-\hat{H})$ represented in the Krylov basis is given by the tridiagonal matrix $T_m$.
We evaluate the determinants recursively by using subdeterminants.
With the definition det$(A_1) = $~det$(M_1)$, we write
\begin{align*}
    x_0 = \frac{1}{z - \alpha_1 - \beta_2^2 \frac{\mbox{\small{det}} A_2}{\mbox{\small{det}} A_1}}
\end{align*}
and fix the subleading determinant by
\begin{align*}
    \frac{\mbox{det}A_{i+1}}{\mbox{det}A_i} = \frac{1}{z - \alpha_{i+1} - \beta_{i+2}^2 \frac{\mbox{\small{det}} A_{i+2}}{\mbox{\small{det}} A_{i+1}}} \,.
\end{align*}
This leads to the continued fraction decomposition 
\begin{align*}
    x_0 = \frac{1}{z - \alpha_1 - \frac{\beta_2^2}{z - \alpha_2 - \frac{\beta_3^2}{z - \alpha_3 - \dots}}} \,,
\end{align*}
which determines the spectral function according to $I(\omega) = -\mbox{Im}\{x_0\}/\pi$.
The spectral decomposition of an initial state $\ket{\psi}$ is defined by the set of spectral functions $I_k(\omega)$, where $k$ labels the individual Fourier components of $\ket{\psi}$.

%\bibliography{SpinLadder_References}

%\bibliographystyle{plain}
\end{document}